\documentclass[12pt]{article}
\usepackage{epsfig,cite,comment}
\newlength{\picwi}
\usepackage[pagewise]{lineno}

\newlength{\twofig}  
\begin{document}
%
\def\nopar{\noindent}
\newcommand{\bec}{Bose-Einstein Correlations}
\def\Ecm{\ensuremath{\sqrt{s}}}
\def\epem{\ensuremath{\mathrm{e}^+\mathrm{e}^-}}
\def\qq{\ensuremath{\mathrm{q\overline{q}}}}
\def\lnu{\ensuremath{\ell\overline{\nu}_{\ell}}}
\def\Zz{\ensuremath{{\mathrm{Z}^0}}}
\def\Zqq{\ensuremath{(\Zz/\gamma)^{*}\rightarrow\qq}}
\def\rWW{\ensuremath{\mathrm W{\mathrm W}}}
\def\rW{\ensuremath{\mathrm W}}
\def\WW{\ensuremath{\mathrm{W}^+\mathrm{W}^-}}
\def\WWqqqq{\ensuremath{\WW\rightarrow\qq\qq}}
\def\WWqqln{\ensuremath{\WW\rightarrow\qq\lnu}}
\def\eeWW{\epem$\to$\WW}
\def\zg4{\Zz/$\gamma^* \rightarrow$}
\def\fig{Fig}
\def\sect{Sect.}
\def\col{Collaboration}
\def\ea{{\it et al.}}
\def\ct{\cite}
\def\lss{\pm\,\pm}
\def\pythia{{\sc Pythia}}
\def\figcap{The systematic uncertainties
for the first two points are shown
as outer vertical bars.}
\def\figcapb{The error bars show 
the statistical and systematic 
uncertainties 
added in quadrature.}
\begin{titlepage}
\flushbottom
\begin{center}
{\large\bf  EUROPEAN 
ORGANIZATION FOR 
NUCLEAR RESEARCH}
\end{center}
\begin{flushright}
{\large  CERN-PH-EP/2004-008}\\
{\large  24th March 2004}\\
\end{flushright}
\vspace*{.8cm}

\bigskip
\bigskip
\begin{center}
\begin{boldmath}
\bf \Huge 
Study of Bose-Einstein Correlations in $\epem\to\WW$ Events at LEP
\end{boldmath}
\end{center}
\bigskip
\bigskip
\begin{center}
{\LARGE\bf The OPAL Collaboration}
\end{center}
\bigskip
\bigskip
\bigskip
\begin{abstract}
\nopar
Bose-Einstein correlations between like-sign charged-particle pairs
in  \eeWW\  events recorded
with the OPAL detector at LEP at centre-of-mass energies between 183 
GeV
and 209 GeV
are studied. 
Recently proposed methods which allow direct searches for 
correlations  
in the
data via distributions of test variables are used to investigate 
the presence of correlations between hadrons originating 
from different W bosons 
in 
\WWqqqq\ events. 
Within the statistics of the data sample
no 
evidence for  
inter--WW Bose-Einstein correlations
is obtained.
The data are also compared with predictions of a  recent implementation of 
Bose-Einstein correlation
effects in the Monte Carlo model \pythia.
\end{abstract}
\vspace{1.cm}
{\centerline {\large \it Submitted to European Physical Journal C}}
\end{titlepage}
\newpage

\begin{center}{
G.\thinspace Abbiendi$^{  2}$,
C.\thinspace Ainsley$^{  5}$,
P.F.\thinspace {\AA}kesson$^{  3,  y}$,
G.\thinspace Alexander$^{ 22}$,
J.\thinspace Allison$^{ 16}$,
P.\thinspace Amaral$^{  9}$, 
G.\thinspace Anagnostou$^{  1}$,
K.J.\thinspace Anderson$^{  9}$,
S.\thinspace Asai$^{ 23}$,
D.\thinspace Axen$^{ 27}$
% ...
}\end{center}\bigskip
\bigskip
$^{  1}$School of Physics and Astronomy, University of Birmingham,
Birmingham B15 2TT, UK
\newline
$^{  2}$Dipartimento di Fisica dell' Universit\`a di Bologna and INFN,
I-40126 Bologna, Italy
\newline
$^{  3}$Physikalisches Institut, Universit\"at Bonn,
D-53115 Bonn, Germany
\newline
% ...
$^{  5}$Cavendish Laboratory, Cambridge CB3 0HE, UK
\newline
% ...
$^{  9}$Enrico Fermi Institute and Department of Physics,
University of Chicago, Chicago IL 60637, USA
\newline
% ...
$^{ 16}$Department of Physics, Schuster Laboratory, The University,
Manchester M13 9PL, UK
\newline
% ...
$^{ 22}$Department of Physics and Astronomy, Tel Aviv University,
Tel Aviv 69978, Israel
\newline
$^{ 23}$International Centre for Elementary Particle Physics and
Department of Physics, University of Tokyo, Tokyo 113-0033, and
Kobe University, Kobe 657-8501, Japan
\newline
% ...
$^{ 27}$University of British Columbia, Department of Physics,
Vancouver BC V6T 1Z1, Canada
\newline
% ...
$^{  y}$ now at CERN
\newline
% ...

\newpage
\section{Introduction}

Bose-Einstein correlations (BEC) between identical bosons are a 
well-known phenomenon in high energy physics~\ct{BECinHEP}. BEC are often 
considered to be 
the analogue of
the Hanbury Brown and Twiss effect~\ct{HBT} in astronomy, 
describing
the interference of identical bosons emitted incoherently. 
However, alternative 
models 
exist such as that
 proposed by B.~Andersson et al.~\cite{Bo_Andersson}, which includes
a coherent particle production mechanism in the framework of
the Lund string model \ct{Bo_Andersson_Book}. 

BEC lead to an enhancement of the production of identical bosons close in
phase space. First reported for pairs of charged pions produced  in
hadron-hadron collisions \cite{GGLP}, 
BEC have been studied for systems of two or more identical bosons 
produced in
various types of collisions, and in particular
in  hadronic \Zz\
decays 
 from \epem\ annihilation
at LEP (see
\cite{Kittel,opal-gen,GideonA,WesM} and
references therein). 

At LEP, BEC have been unambiguously established between the particles 
originating
from one 
hadronically decaying W, representing so-called {\it intra}--W BEC
\cite{lep-wbec,l3-ww}. 
The aim of this paper is to search for
evidence of
BEC between the particles originating from {\it different} W bosons,
i.e. for
 {\it inter}--WW BEC in \eeWW\ events. A recent L3 study \cite{l3-ww} 
using the same method \cite{CdWK} as we use here 
shows no  evidence for
inter--WW BEC. 

Two hadronically decaying W bosons provide a unique opportunity to study
two partially 
overlapping hadronic systems allowing this important aspect of 
BEC to be explored \cite{Eddi}. 
The typical separation of the two W decay vertices in
\WWqqqq\ events is of 
the order of 0.1~fm, while the hadronization scale is of the order of 
a few
fm. 
In  incoherent scenarios, the difference between the correlations inside
the 
hadronic system of one W and the correlations between the two hadronic systems 
depends on the overlap region of these two  systems. 
In a coherent scenario, the correlations between the two systems 
may not exist at all, and the two systems would then decay independently 
provided
there is no 
colour flow between them.

Inter--WW BEC effects (along with colour
reconnection effects) are among   the largest 
uncertainties in the 
determination of the W mass in the \WWqqqq\ channel at LEP 
\cite{lep-rep,ewwg}. 
If inter--WW BEC affect
particles 
from different W bosons, 
this
can 
disturb the
W mass 
determination from the \qq\ invariant masses. Initial 
predictions of various 
Monte Carlo (MC) models gave an uncertainty of up to 100 MeV
on the W mass
arising from BEC.
Excluding some of the more extreme models, the most 
recent LEP estimate for the uncertainty is now 35~MeV \ct{ewwg}.

\section{Analysis method}
\label{analysis}

BEC are usually presented in terms of two-particle densities, 
$\rho_2(Q)$, measured as

\begin{equation}
\rho_2(Q) =
\frac{1}{N_{\rm events}}\frac{\mathrm{d}N_{\rm pairs}}{\mathrm{d}Q}\:,
\label{rho2}
\end{equation}

\nopar
for the number $N_{\rm pairs}$
of pairs 
of identical bosons with four-momenta $p_1$
and
$p_2$ and
$Q=\sqrt{- (p_1 - p_2)^2}$ 
in the number $N_{\rm events}$ of events under study.  
 The correlations can be expressed in terms of the normalised 
inclusive
two-particle density,

\begin{equation}
R_2(p_1,p_2) = 
\frac{\rho_2(p_1,p_2)}{\rho_1(p_1)\rho_1(p_2)}
\:, 
\label{r2}
\end{equation}

\nopar
i.e. the ratio of the 
two-particle density $\rho_2(p_1, p_2)$, 
usually 
measured 
as a 
function of $Q$, 
$\rho_2(Q)$
(Eq.~(\ref{rho2})),
 to the product of the two single-particle densities 
$\rho_1(p_1)$ and $\rho_1(p_2)$. The single-particle density $\rho_1(p)$ 
is measured as $\rho_1(p)= 1/N_{\rm 
events}\cdot {\rm d}n_{\rm ch}/{\rm d}p$, where $n_{\rm ch}$ is
the multiplicity 
of
charged particles. 

In this paper
we use the method proposed
in~\ct{CdWK} 
 to study inter--WW BEC. This method  
 allows a direct search for inter--WW correlations from the data, with no 
need of MC models. 
If the two W bosons decay 
independently, then  the two-particle  density $\rho_2^{\rWW}(p_1, p_2)$ 
in \WWqqqq\ events
can be written as
 the sum of the  two-particle densities $\rho_2^{\rW^{+(-)}}(p_1, p_2)$ of 
the individual $\rW^{+(-)}$ and an additional part consisting of the 
product of   
single-particle densities,  $\rho_1^{\rW}(p_1)$ and  $\rho_1^{\rW}(p_2)$, 
from different W bosons:    

\begin{equation}
\rho_2^{\rWW}(p_1, p_2) = 
\rho_2^{\rW^+}(p_1, p_2) + \rho_2^{\rW^-}(p_1, p_2) +
\rho_1^{\rW^+}(p_1)\rho_1^{\rW^-}(p_2) +
\rho_1^{\rW^-}(p_1)\rho_1^{\rW^+}(p_2).
\label{aaa}
\end{equation}

\nopar
For pairs of charged hadrons, symmetry arguments imply that the 
two-particle density  of W$^+$, constructed from pairs of negatively 
charged particles 
added to that constructed from pairs of positively charged 
particles, is identical to 
that of W$^-$; thus  
Eq.~(\ref{aaa}) becomes

\begin{equation}
\rho_2^{\rWW}(p_1, p_2) = 2\, \rho_2^{\rW}(p_1, p_2) + 
                        2 \, \rho_1^{\rW}(p_1) \rho_1^{\rW}(p_2).
\label{rho2WW}
\end{equation}

\nopar
The two-particle densities $\rho_2^{\rWW}$ and $\rho_2^{\rW}$
are
determined from  \WWqqqq\ events and  \WWqqln\
events
respectively. For the latter the lepton or its decay products
 are removed from the event.

The product of single-particle densities,
$\rho_1^{\rW}(p_1) \rho_1^{\rW}(p_2)$, is determined 
by
constructing artificially ``mixed'' \WWqqqq~events
from the hadronic decay products of two  $\WW\to \qq(\lnu)$ events,  
as described in \sect~\ref{section:mix}.
The charge of the lepton is used to determine the charge of the 
hadronically decaying W system when constructing these events. 
Pairs of particles originating from different W bosons in the mixed events 
are uncorrelated by construction; the two-particle density formed from 
such pairs is termed $\rho_{\rm mix}^{\rm WW}(p_1,p_2)$.
After integration over all momenta, but keeping $Q$ fixed,   
Eq.~(\ref{rho2WW}) reads
\begin{equation}
\rho_2^{\rWW}(Q) = 2\,\rho_2^{\rW}(Q) + 2\,\rho_{\rm mix}^{\rm
WW}(Q)\,.
\label{rho2WWinQ}
\end{equation}

The presence or absence of BEC between particles from different W
bosons in \WWqqqq~events can be tested by verifying the 
  equality
between the two sides of Eq.~(\ref{rho2WWinQ}) using distributions
of different test variables to be defined below. 
This allows 
a variety of possibilities 
to be explored 
in an
experimental search for inter--WW BEC.
Selecting the
 variables
which are most sensitive to inter--WW BEC, a method can then be
devised which is best suited to evaluate the systematic error on
the measurement of the W mass at LEP caused by possible inter--WW
BEC.

We start by studying the distribution $\Delta\rho(Q)$, which probes
the independent hadronic decay of the two W bosons by comparing the
two-particle densities from fully hadronic events
(where all
 possible correlations are present) with the two-particle densities
 of artificially constructed events containing only intra--W
 correlations,
 \begin{equation}
 \Delta \rho(Q) = \rho_2^{\rWW}(Q) - 2\,\rho_2^{\rW}(Q) -
 2\,\rho_{\rm mix}^{\rm WW}(Q)\,. 
\label{deltarho}
 \end{equation}
 
 \nopar We also consider the integral of the $\Delta \rho(Q)$
 distribution, integrated
from $0$ to $Q_{\rm max}$:
\begin{equation}
J\equiv \int_0^{Q_{\rm max}} \Delta \rho(Q) \mathrm{d}Q\,,
\label{integral}
\end{equation}

\nopar 
where bin-to-bin statistical fluctuations in the 
$\Delta\rho(Q)$ 
distributions are reduced.

\nopar
In addition, we study,
as a direct measure of genuine
inter--WW correlations \ct{Eddi}, 
the inter-source correlation function, 
\begin{equation}   
\delta_I (Q) = \Delta \rho(Q) /  \rho_{\rm mix}^{\rWW}(Q)\,,
\label{deltai}  
\end{equation}

\nopar
and the $D$-ratio \ct{l3-ww}, 
\begin{equation}   
D(Q) = \frac{\rho_2^{\rWW}(Q)}{ 2\,\rho_2^{\rW}(Q) + 2\,\rho_{\rm 
mix}^{\rm
WW}(Q)}\,
\label{bigd}  
\end{equation}

\nopar 
in which contamination from semileptonic events and artificial effects  
due to the 
mixing procedure
are 
expected to be reduced.

To disentangle the BEC effects from other possible correlation
sources (such as energy-momentum conservation or colour
reconnection), which are supposed to be the same for like-sign
($\lss$) and unlike-sign $(+\,-)$ charge pairs, we analyse the
double difference,
\begin{equation}
\delta \rho (Q)= \Delta \rho(\lss) -  \Delta \rho(+\,-)\,,
\label{deltar}
\end{equation}

\nopar
its
corresponding integral according to  Eq.~(\ref{integral}),  
as well as 
the 
inter-source 
correlation 
functions 
difference,
\begin{equation}
\Delta_I (Q)= \delta_I(\lss) -  \delta_I (+\,-)\,,
\label{deltaii}
\end{equation}

\nopar
and  the double ratio, 
\begin{equation}
d(Q)= D(\lss)/D(+\,-)\,.
\label{smalld}
\end{equation}

\nopar

In Eqs.~(\ref{deltar}), (\ref{deltaii}) and (\ref{smalld}),
 contributions from correlations other than BEC are expected to
 cancel, thus only BEC effects will affect these distributions.
 Moreover, any
potential bias introduced by imperfections in the event mixing
procedure
  should be strongly reduced. The distributions $\delta
 \rho(Q)$, $\Delta_I$ and $d(Q)$ have the advantage of giving
access to inter--WW BEC directly from data and do not rely on
Monte Carlo modelling.

Another distribution which corrects for detector effects and
possible imperfections in the event mixing procedure, but which
introduces a  MC model dependence, has been advocated
in \ct{l3-ww}. The  double ratio $D'$ is defined as
 \begin{equation}
D'(Q) = \frac{D(Q)}{D_{\rm no-BEC\: MC}(Q)}\,,
\label{dprime}
\end{equation}

\nopar
where $D_{\rm no-BEC\: MC}(Q)$ is obtained from a Monte Carlo
simulation without BEC but which includes 
other 
 possible
correlations.

If there are no correlations between particles originating from
different W bosons, the variables defined above will,
 by construction, 
have the values:
$\Delta \rho(Q) = \delta \rho(Q) = 0$ and $D(Q) = D'(Q) = d(Q) = 1$ 
for
all $Q$. 
The inter--WW function $\delta_I(Q)$, Eq.~(\ref{deltai}), can have
arbitrary
(positive or negative)
values in the case 
where
the W bosons decay  products 
overlap only partially in momentum space. For fully overlapping
and uncorrelated WW decays $\delta_I\equiv0$\ct{Eddi}.

\section{Experimental details}
\label{expdet}

\subsection{The OPAL detector}
\label{sec-det}

The OPAL detector has been described in detail
elsewhere~\ct{detector}. The analysis presented here relies mainly on
the  
charged particle trajectories
reconstructed using a set of cylindrical central tracking detectors 
within a solenoid that provides an axial magnetic field of 0.435~T\@.
Electromagnetic energy is measured by a
lead-glass calorimeter located outside the magnet coil.
The innermost tracking detector is a
silicon microvertex detector, which consists of two layers of silicon 
strip 
detectors, allowing at least one hit per charged particle track to be 
measured in the 
angular region 
$|\cos\theta|<0.93.$\footnote{The OPAL right-handed 
coordinate system is defined
such that the origin is at the geometric centre of the jet chamber, $z$ is 
parallel to, and has positive sense along, the  e$^-$ beam direction, $r$ is 
the coordinate normal to $z$, 
$\theta$ is the polar angle with respect to 
+$z$ and $\phi$ is the azimuthal angle around $z$.}
 It is surrounded by the vertex 
drift 
chamber and the jet chamber, which is about 400~cm in length and 185~cm in 
radius, and provides up to 159 space points per track and also measures 
the
ionization energy loss of charged particles. 
The 
$z$-chambers,
which considerably 
improve the measurement of the trajectories in $\theta$, complement the 
tracking system. The combination of these chambers leads to a 
transverse momentum
resolution of 
$\sigma_{p_{\rm t}}/p_{\rm t}= \sqrt{(0.02)^2+(0.0015\,  
p_{\rm t}/{\rm GeV})^2}$.
Track finding is nearly 100\% efficient within the angular region 
$|\cos \theta |<0.92$.
 The experimental $Q$ resolution, $\sigma_Q$, is directly
related to that of $M$, the invariant mass of the particle 
pair: $M^2=Q^2+4m_{\pi}^2$. For  $\pi^{+} \pi^{-}$ pairs from
$\mathrm{K^{0}_{S}}$ decays, the mass resolution is found to be
$\sigma_M = 7.2\pm 0.1$~MeV \ct{ksks_1}, implying that, at
$Q=0.41$~GeV which is typical of the region of $Q$ affected by BEC,
$\sigma_Q=8.7$~MeV. For all distributions presented here, a bin
size of 40~MeV is used, much larger than the experimental
resolution in the region of interest.

\subsection{Track and event selections}
\label{selections}

This study is carried out using data taken at \epem\ centre-of-mass
energies  $\sqrt{s}$ between 183 and 209~GeV  with an integrated 
luminosity of 
approximately  680~pb$^{-1}$. 
For the charged particles used in the BEC analysis the number of recorded
hits in the jet chamber is required to be 
at least 40 and
larger than 50\% of the expected 
number at the given $\cos\theta$.
Tracks must have a momentum component in the plane perpendicular to the
beam axis of greater than 0.15~GeV, and a measured momentum of less
than 100~GeV. In addition, they are required to 
have a good $\chi^{2}$
per degree-of-freedom 
  for the track fits in the planes perpendicular and parallel to
the beam direction.  The extrapolated point of closest approach of each
track to the
collision axis
is required to be less than 2~cm in the $r\phi$-plane and less 
than 
25~cm in $z$. The selected particles are assumed to be charged pions.

Two mutually exclusive event samples are 
selected: 
the {\em fully hadronic} event sample, 
$\WW$ $\rightarrow\qq\qq$,
where both W bosons decay
hadronically and the {\em semileptonic} event sample, \WWqqln,
where one W decays hadronically and the other  leptonically.
Both selections are described in detail in~\ct{event_selections}. 
 The fully hadronic selection uses a likelihood weight
${\cal{L}}$ based  on a set of variables which characterize the \WWqqqq\ 
decays. To suppress  \zg4~jets
background, in 
this
 analysis the requirement on this  
likelihood weight is tightened from the 
standard value of ${\cal{L}}>0.23$ to  ${\cal{L}}>0.55$.
This reduces the residual  \zg4 background from 
15\% to  8\% of selected events, 
whilst  
reducing the signal efficiency e.g. at $\sqrt{s}= 189$~GeV from  86\% to 
71\%.

The numbers of \WWqqqq\ 
events selected in the data are 1721 for 
$\sqrt{s}=$~183--192~GeV, 1290  for $\sqrt{s}=$~196--200~GeV and 1459 for 
$\sqrt{s}>$~202~GeV.
The corresponding numbers in 
the  channel \WWqqln\ are   
1720, 1300 and 1513.
This channel includes the $\WW\to\qq{\rm e}{\overline \nu}_{\rm e}$ and 
$\WW\to 
\qq\mu 
{\overline \nu}_{\mu}$ 
events, 
and those $\WW\to \qq\tau {\overline \nu}_{\tau}$ events where the
$\tau$ 
lepton 
decays 
to e, $\mu$ or one charged hadron.      
 The  fraction of selected 
background of \zg4~jets 
in the 
\WWqqqq\ channel 
 is 
almost independent of the centre-of-mass energy.
 Events of the type  $\Zz\Zz \to$~jets are not considered as background
 since the ZZ
 system should be affected 
by BEC effects in a similar way to the WW signal.
Although no 
correction is made for a few percent background
contribution in the \WWqqln\ channel, this background was 
taken into account 
 in assessing
 the systematic uncertainties.

The treatment of the   \zg4~four jets
background 
in the 
\WWqqqq\ channel 
requires special attention, because these background 
events could mimic the signal of inter--WW BEC. This will be 
further discussed 
below. 

  All 
data and Monte Carlo
distributions presented in this paper 
are at the {\it detector level}, i.e. they are
not corrected for effects of detector acceptance and resolution. Background 
contributions have been subtracted from the data.

\vspace*{0.6cm}

\section{Monte Carlo modeling of BEC}
\label{MonteCarlo}

Throughout this analysis
the  {\pythia} 6.1 Monte Carlo program  \ct{bib-jetset} is used to 
demonstrate  
the sensitivity of the analysis to BEC effects for several
different scenarios.
Monte Carlo samples of 
  about 30
times the number of data events are 
generated 
at energies of 189~GeV, 
200~GeV 
and 
206~GeV,
and processed through a full simulation of the detector \cite{osim}.

In  \pythia, BEC effects are 
implemented via  
the  {\tt PYBOEI}~\ct{PYBOEI} model (the option BE$_{32}$ was used here).
In the model, the particle momenta are adjusted 
to produce 
a BE enhancement of the form
\begin{equation}
R_2(Q) \sim 1+\lambda\cdot\mathrm{exp}(-r^2Q^2)\,,
\label{Eq:1}
\end{equation}

\nopar
where, in the so-called static 
incoherent 
picture, 
$r$ represents
 the source radius and $\lambda$ the BEC ``strength''.
Global energy-momentum conservation is achieved by also adjusting 
the momenta of
particles in 
unlike-sign 
pairs.
Various  implementations of the model 
can be tested, among them the full
(intra--W
plus  
inter--WW) 
BEC, the  intra--W BEC (no inter--WW BEC), and the no--BEC options.

In the present analysis, a
Gaussian parameterisation with the {\tt PYBOEI}
parameters 
{\tt PARJ(92)} ($\equiv \lambda$) = 2.15 and {\tt PARJ(93)} = 0.25~GeV
(which leads to $r=0.73$~fm) 
is used for the full--BEC and intra--W BEC only  cases.
The   \pythia\ QCD and  fragmentation parameters, based on a previous 
OPAL tune \ct{opal-tune}, along
with the  {\tt PYBOEI} BEC model parameters 
have been  retuned together to the  \Zz\  data.{\footnote 
{Only the main  QCD/fragmentation parameters change
with respect to the OPAL standard parameters:
 {\tt PARJ(81)} $(\equiv \Lambda_{\rm QCD}) = 0.25\to 0.27$~GeV,
 {\tt PARJ(82)} $(\equiv  Q_0) = 1.90  \to 1.75$~GeV, 
 {\tt PARJ(42)} (Lund $b$ parameter) $= 0.52 \to 0.48$~GeV,
 {\tt PARJ(21)} $(\equiv \sigma_{pt}) = 0.40 \to 0.45$~GeV.}}
With these parameters, the two-particle distribution of the \Zz\ data,
with both data and MC 
normalised to a MC without BEC, is described to better than 2\% in the
$Q$ range
between 0.05 and 0.6 GeV, 
as shown in \fig.~\ref{Fig:tune}. 
 For $0<Q<50$~MeV, the MC fails to describe the data as a result of
artificial effects of the implementation of the correlations in
the model. For intermediate values of  $Q$,  up to about 0.5~GeV,
which is the region of interest for the study of BEC, the
agreement between data and MC is very good. A small discrepancy
for higher $Q$ values is seen. This can be explained from the fact
that the full integral over $Q$
is related to the mean number of particle pairs. Since the {\tt PYBOEI}
model does not change 
the 
event multiplicity distribution, any
difference in the $Q$-distributions at small $Q$ has to be
compensated elsewhere.

 The MC tuned on inclusive \Zz\ decays at LEP1 overestimates BEC 
in \Zz\
 events with jet topologies similar to the \WWqqqq\ topology. The
two-particle density of the BEC Monte Carlo used to subtract the
\zg4\ background in the \WWqqqq\ events is therefore corrected
bin-by-bin by
applying a scaling factor of
the ratio of the two-particle densities
of data and MC for multi-jet events at $\sqrt{s}=$~91~GeV. Because
the \WWqqqq\ selection \ct{event_selections} is mainly based on
variables scaled to the centre-of-mass energy, it is also suitable
for the selection of \Zz\ events.

\section{Event mixing technique}
\label{section:mix}

To measure the quantities defined in Eqs. 
(\ref{deltarho})--(\ref{dprime}), the
two-particle density $\rho_{\rm mix}^{\rWW}(Q)$ needs to be determined.
This is achieved by mixing the hadronic parts of 
two semileptonic \WWqqln\ 
data 
events, after removing the leptonic parts 
of the events. 
By combining two hadronic W
decays recorded in different events, an artificial event can be
constructed which is guaranteed to have no inter--WW correlations.
The particles originating from one hadronically decaying W are fixed, 
while the particles from the second W are rotated 
in azimuth
such that the two 
W bosons are back-to-back in $\phi$.
The 
 mixed event has a topology similar to that of a real \WWqqqq\ event.
The two decaying 
W bosons are selected to have opposite charge, and the centre-of-mass 
energies of 
the two 
semileptonic events are required to be similar, such that
\begin{equation}
\left|E_{\rm cms}^{\rm event\,1} - E_{\rm cms}^{\rm event\,2}\right| \le
5~\mathrm{GeV}.
\label{edif}
\end{equation}

\nopar
The direction of the hadronically decaying W is determined to a precision 
of 80~mrad in the polar angle $\theta$
using 
the sum of momenta of  all charged particles 
and  
clusters  of energy in the electromagnetic calorimeter   
 which are not associated with 
tracks.
The difference in the reconstructed 
$\theta$ angle  between the two hadronic 
W bosons is required to be 
\begin{equation}
\left|\theta_{\rW^+} - \theta_{\rW^-}\right| \le 75\,\, {\rm mrad}\qquad 
{\rm or}\qquad
\left|(\pi-\theta_{\rW^+}) - \theta_{\rW^-}\right| \le 75\,\,
\mathrm{mrad}.
\end{equation}

\nopar
This ensures that both W bosons were 
 originally oriented towards 
detector regions 
which have the same track detection properties, 
either 
in  the same hemisphere or in 
opposite ones.
In the 
 former
 case all charged particle momenta 
 of one W 
 are    reflected into 
the opposite hemisphere.
The mixed \WWqqqq~events are then passed through the 
regular \WWqqqq~event selection, which  rejects 26\% of all mixed events.

To check that the mixed events resemble closely real \WWqqqq\
events, the event shape variables and single-particle spectra are
compared. Good agreement of the latter is essential, since the
mixing term in Eq.~(\ref{rho2WW}) depends directly on these
spectra. Fig.~\ref{Fig:spectra} shows the distributions of
rapidity, transverse momentum and $\Phi$, the angle between the
particle direction and a plane determined by the incoming beam
direction and the thrust axis, for real and mixed events after the
event selection. The thrust axis of the mixed events is used as
the reference axis for the single-particle spectra. Good agreement
is observed between the distributions for mixed WW events and
those of real WW events. As a further check, thrust, oblateness
and aplanarity \cite {opal-tune} distributions are compared in
\fig.~\ref{Fig:eventshapes}. The distributions of mixed events
agree reasonably well with those for real WW events. The thrust
distributions show some differences which, however, are due to
event selection effects and are unimportant for studies of BEC.

The two-particle distributions for the \WWqqqq\ events are scaled by 
a small factor ($\approx 1.04$) to correct for the slightly different 
track 
selection efficiencies for the \WWqqqq, \WWqqln\ and the 
constructed mixed \WWqqqq\ events.

\section{Systematic uncertainties}
\label{systs}

The data discussed in the next section will be shown with 
statistical and systematic uncertainties
 added in quadrature.
All systematic errors are dealt with on a bin-by-bin basis by calculating
the effect of each systematic on each bin separately, and summing all the
variations for a bin in quadrature. 
Several  sources of systematic effects  have been studied.

\begin{itemize}

\item Track selection: We have repeated
the study  requiring 
the measured distance of closest approach of each track  to the 
 collision axis 
to 
be
 less than 1~cm (rather than 2~cm) in the $r\phi$-plane. Alternatively,
an additional requirement that the mean energy loss ${\rm d}E/{\rm d}x$ 
value is compatible with the pion hypothesis at 99\% confidence level was 
imposed.
These two changes, giving a systematic uncertainty in the test 
distributions 
of
less than 2\% and 6\%, respectively, have been summed in
quadrature to give the total systematic uncertainty due to charged track 
quality
selection criteria.

\item Background in the \WWqqln\ channel: We have repeated our 
analysis
using only 
$\WW\to\qq{\rm e}{\overline \nu}_{\rm e}$ and
$\WW\to \qq\mu   {\overline \nu}_{\mu}$
 semileptonic 
events which are selected with a high purity. Final states selected as 
$\WW\to \qq\tau {\overline \nu}_{\tau}$
events were removed from the semileptonic two-particle densities 
and were not used for the event mixing. 
This gives a systematic uncertainty of less than 3\% on the test 
distributions.

\item
 Event selection for \WWqqqq:
 We 
have repeated the analysis with  the OPAL standard likelihood weight 
requirement of
$\cal L>$  0.23 instead of $\cal L>$  0.55 used 
here. 
This change introduces a systematic uncertainty of less than 5\%
on
the
variables studied.

\item 
Monte Carlo correction of the \zg4~four jets background:
We have repeated the analysis subtracting \Zz\ background  which was 
not 
scaled due to the topology difference as explained in 
\sect~\ref{MonteCarlo}.
 This introduces a systematic uncertainty of less than  4\% on the test 
distributions. 

\item Event mixing procedure: We have repeated the analysis 
modifying the  main event mixing criteria. 
We varied the
requirement on the reconstructed
$\theta$ angle in the range between 50 and 100 mrad.
This gives a systematic uncertainty of less than 1\% in
both cases. 
Changing  the requirement on  the energy difference
of the mixed W bosons, Eq.~(\ref{edif}),
 from 5 to 8~GeV 
results in a difference of  less than 3\% for the distributions under 
study.

\end{itemize}

In addition, the effect of colour reconnection was studied. Using the 
implementation of this effect in the {\sc
Ariadne} model AR2 and AR3 \ct{ar-cr}, no 
 significant influence on the 
results presented
here was found
after the \WWqqqq\ two-particle
distributions are scaled to have the same 
 mean
particle
pair multiplicity as 
 that of 
the 
mixed 
events.

\section{Results}
\label{results}

Figure \ref{Fig:qdist_lus}  shows \pythia\ MC predictions
 for the two-particle density, defined in Eq.~(\ref{rho2}), of
like-sign and unlike-sign particle pairs for \WWqqqq, \WWqqln\ and
mixed events, and for the \zg4~four jets background sample,
calculated for the three BEC scenarios: the full--BEC scenario where a 
low-$Q$
Bose-Einstein enhancement is simulated for all like-sign particle
pairs, including those where the two hadrons originate from
different W bosons; the intra--W BEC scenario where the BEC effect acts 
only on
particles originating from the same W, and the no--BEC scenario,
where the BEC effect is not simulated.

Background events from \zg4~jets decays  which satisfy the WW
selection criteria have a higher multiplicity than \WWqqqq\
events.  As \fig.~\ref{Fig:qdist_lus} shows,  $\rho_2(Q)$ for such
events is large compared to that of real \WW\ events. Although the
fraction of background is low, a careful subtraction is necessary.
In the present analysis, the \zg4~jets contribution is subtracted
bin-by-bin from the data using the MC predictions. 
 Note that this is the only instance where a MC model-dependence
enters the analysis; all other information needed is derived
directly from the data themselves.

To determine the statistical errors on the distributions shown
 in the following, a statistical
sampling technique has been used instead of the conventional
method of error propagation. For each individual distribution, the
effect of statistical fluctuations is simulated by randomly
sampling the content of each bin,
using the full covariance matrix to account for bin-to-bin
correlations. The means of the distributions are set equal to the
observed bin contents. The statistical errors on $J$,
Eq.~(\ref{integral}), and of each bin of the distributions
studied,  are then estimated from the dispersion of the results,
after repeated sampling of the input distributions. A re-sampling
frequency of 1000 was used for this analysis.

The data results presented hereafter are compared to the predictions of
the three BEC MC scenarios described above. This allows the assessment of
the experimental sensitivity of the various test distributions, defined in
Sect.~\ref{analysis}, to inter--WW correlations of the type and 
strength considered in \pythia.

Figs.~\ref{Fig:drhos}(a) and \ref{Fig:drhos}(b)  show the
experimental  $\Delta \rho(Q)$-distributions, defined in
Eq.~(\ref{deltarho}), for like-sign and unlike-sign pairs.
 Both $\Delta\rho({\lss})$ and $\Delta\rho({+\,-})$ are, for all $Q$,
 compatible with zero within uncertainties, as is expected if the
 W$^+$ and W$^-$ decay independently. In particular, $\Delta\rho({\lss})$
 shows no evidence for a strong inter--WW BEC effect.

The intra--W and no--BEC MC predictions are zero within errors. This 
shows
that the event mixing technique used is adequate and does not introduce
strong methodological biases. 
 For the full--BEC scenario,  the inter--WW BEC effect is clearly
visible for like-sign particle pairs at low $Q$, $0.04 <Q<
0.48$~GeV, \fig.~\ref{Fig:drhos}(a). Note, however, that a small
and rather broad enhancement of the full--BEC curve over the other
BEC scenarios is predicted also for unlike-sign particle pairs,
\fig.~\ref{Fig:drhos}(b). These artificial correlations arise from the way
energy-momentum conservation is locally enforced in {\tt PYBOEI} and
affects all particles, whatever their charge.
 
Whereas the result for $\Delta\rho({\lss})$ is, by itself,
consistent with the hypothesis of no inter--WW BEC, comparison with
the full--BEC predictions leads to the conclusion that the
experimental sensitivity is insufficient to be able to exclude
inter--WW correlations of the type and size as introduced in the
BEC model of \pythia.

In  \fig.~\ref{Fig:int}, 
the integrated $\Delta\rho$ distributions, $J(\lss)$ and $J(+\,-)$,
Eq.~(\ref{integral}), computed by summation over bins up to
$Q_{\rm max}$, are shown. The data for
$J(\lss)$ are consistent with zero, as are the intra--W  and
no--BEC predictions. 
 As expected, significantly larger values are observed
                    for the model with inter--WW BEC. Positive values are
                    also predicted for $J(+\,-)$; these result from the 
broad 
                    enhancement at low $Q$ observed in 
\fig.~\ref{Fig:drhos}(b).

The values of $J(\lss)$ for $Q_{\rm max}=0.48$~GeV are given in the first
       row of Table 
\ref{tab-fit}. The experimental value is consistent
                    with that for the intra--W and no--BEC scenarios, but 
                    differs from that for full BEC effects by 2.2 
                    standard deviations.

In \fig.~\ref{Fig:deltaI}, we present the inter--WW correlation
function $\delta_I$, Eq.~(\ref{deltai}), for like-sign and
unlike-sign pairs. In data and for all models except the full--BEC
case, $\delta_I(\lss)$ is consistent with zero even in the low-$Q$
region, with the largest deviation in the second lowest data
point.  Although the data do not show 
 a significant signal of
 inter--WW BEC,
 the low data  statistics do not allow 
 the different scenarios
 to be distinguished.

Figs.~\ref{Fig:D}
and
\ref{Fig:Dprime}
 show the
distributions $D(Q)$, Eq.~(\ref{bigd}), and
 $D'(Q)$,
Eq.~(\ref{dprime}), respectively, for like-sign and unlike-sign
particle pairs. The data points are compatible with  unity for
all $Q$, as expected for independent W$^+$ and W$^-$ decays. The
same holds for the intra--W and no--BEC scenarios.
 The  data are  consistent with both the full 
 and
 intra--W BEC scenarios as predicted by \pythia.

The distributions $D(\lss)$ and $D'(\lss)$ (and $d(Q)$, see
below) have been fitted with an empirical
parametrization~\ct{Eddi} of the form

\begin{equation}
f(Q) = N (1 + \delta \cdot Q) (1 + \Lambda\cdot\exp(-Q/R)).
\label{fit}
\end{equation}
 The fits were performed in the interval $0.04 < Q< 2$~GeV using
the full covariance matrix of the corresponding distributions.
 In Eq.(\ref{fit}), $N$, $\delta$, $\Lambda$ and $R$ are fit
parameters: $N$ is the overall normalisation, $\delta$ takes into
account effects due to potential long-range correlations,
$\Lambda$ and $R$ are, respectively, a measure of the ``strength''
and  width of the enhancement expected from inter--WW BEC. 
 Since the predictions of the MC scenarios without inter--WW BEC and the 
data are compatible with a constant value of $D(Q)$, $R$ was first 
determined 
from a fit to the respective distribution for a MC event sample with 
full BEC. This reduces the number of free fit parameters. The 
values 
obtained 
are given in Table
\ref{tab-fit}.\footnote{From now on, we use a notation corresponding 
                   to the fitted variable to denote the parameters
                   resulting from the fits.}
In all other fits, $R$ was kept fixed at these
values.

To evaluate the systematic uncertainties, in addition to the effects
listed in \sect~\ref{systs},
  the fits have also been
performed starting at $Q=0$~GeV instead of $Q=0.04$~GeV,
 and  fitting the distribution up to
 $Q=1$~GeV  and $Q=4$~GeV.
In addition, the fits were repeated either changing the values of the 
width parameter
 $R'^{\, \rm full\:\: BEC}$ or $R^{\, \rm full\:\: BEC}$  
within the errors quoted, or omitting\footnote{The corresponding
parameters $R$ used are $R^{\, \rm full\:\:
BEC}=0.210\pm0.022$~GeV and $R'^{\, \rm full\:\:
BEC}=0.184\pm0.031$~GeV.}
 the
factor $1+\delta\cdot Q$ in Eq.~(\ref{fit}). In view of the
deviation between \Zz\ data and the corresponding MC with BEC,
shown in
 \fig.~\ref{Fig:tune}, the fits have been repeated with the interval 
$0.5\leq Q\leq 1$~GeV
 excluded.  In this case only a
negligible change in the results occurred.

The values of the ``strength''  parameter obtained from fits to $D(\lss)$ 
($\Lambda$) and 
$D'(\lss)$ ($\Lambda'$)
for the data and for the MC are collected in Table \ref{tab-fit}
together with the other parameters. The $\Lambda$ values 
are also shown in \fig.~\ref{Fig:fit}.
 The contributions to the systematic uncertainties on $\Lambda$ and 
$\Lambda'$
are
listed in Table \ref{tab-events}. The measured values of $\Lambda$
and $\Lambda'$ 
differ from the full--BEC scenario values
by 1.5  and 
1.3  standard deviations, respectively.

As explained in \sect~\ref{analysis}, the distributions considered so
far are sensitive not only to inter--WW BEC but also to inter--WW
correlations of non-BEC origin. The effect of the latter can be
eliminated or at least reduced by considering the
distributions $\delta\rho$, Eq.~(\ref{deltar}), $\Delta_I$,
Eq.~(\ref{deltaii}), and the $d$-ratio, Eq.~(\ref{smalld}). In
MC calculations, intra--W predictions for the terms
involving unlike-sign pair distributions are used
 to account for the artificial correlations introduced  by {\tt
PYBOEI} discussed earlier. The results are presented in
Figs.~\ref{Fig:drhos}(c), \ref{Fig:deltaI}(c) and \ref{Fig:D}(c).

From \fig.~\ref{Fig:drhos}(c), where the $\delta\rho$ function is 
shown, one can see that there is no
evidence for inter--WW BEC in the
 small $Q$ region: within the
uncertainties, the data points are compatible with zero, as the intra--W 
and
  no--BEC MC predict. 
In addition, the integral of $\delta \rho$, $J(\lss) - J(+\,-)$, 
is
consistent with
zero in the data, with $J(+\,-)=0.39 \pm 0.28({\rm stat}) \pm
0.28({\rm syst})$, calculated up to $Q_{\rm max}=0.48$~GeV. Similar 
results are obtained from the MC
$\delta \rho$-integrals for intra--W  and no--BEC scenarios with
$J(+\,-) =0.24\pm0.10$ and $0.20\pm 0.10$, respectively, while the
$\delta \rho$-integral for the full--BEC scenario is significantly
positive with
$J(+\,-)= 0.68\pm 0.10$.

 The same conclusions can be drawn from
Figs.~\ref{Fig:deltaI}(c) and \fig.~\ref{Fig:D}(c). 
In \fig.~\ref{Fig:deltaI}(c) we show the difference 
$\Delta_I$, Eq.~(\ref{deltaii}), between the inter-source  
like-sign and unlike-sign correlation 
functions, $\delta_I(\lss)$ and $\delta_I(+ -)$, Eq.~({\ref{deltai}); all 
the data points, 
including the first two points, 
are compatible with zero, 
within  the errors. 
In \fig.~\ref{Fig:D}(c) we show the
measured
$d$-ratio, $d(Q)=D(\lss)/D(+\,-)$;  within the
uncertainties, all the data points
are compatible with unity.

The distribution $d(Q)$  has also been fitted with the
parametrization Eq.~(\ref{fit}) in the interval $0.04<Q<2$ GeV.
The parameter $R\equiv R_d^{\, \rm full\: BEC}$
was determined from a fit to $d(Q)$ obtained in the
full--BEC MC; its value is listed in Table \ref{tab-fit}
along with other fitted parameters. The value
for the ``strength'' parameter $\Lambda_d$ is also shown in 
\fig.~\ref{Fig:fit}.
The contributions to the systematic uncertainties on $\Lambda_d$ are
listed in Table \ref{tab-events}. When the factor $1+\delta\cdot
Q$ was omitted, $R_d^{\, \rm full\: BEC}=0.158\pm0.028$ was used in
the fit. The $\Lambda_d$ parameter, which is complementary to that
derived from $D(Q)$ or $D'(Q)$ of like-sign pairs, but with strongly 
reduced
contributions from effects other than BEC, is compatible with zero
for the data. For the MC event sample with full BEC, $\Lambda_d$
deviates from the data value by 2.1 standard deviations.

Several variables have been studied in this analysis
to search for BEC from different W bosons;
it is of interest to identify
the variable and/or distribution which has the largest sensitivity 
to the
inter--WW BEC and can be used for the estimation of the systematic 
errors due
 to BEC on measurements of the W mass.
 We compare the difference between the predictions of the 
two MC scenarios, the 
full--BEC and intra--W 
BEC models, and
 the data, using 
the values of the 
integral  
$J(\lss$) 
and the fitted 
``strength'' parameters given in Table \ref{tab-fit}. 
 The separation power of the
$J(\lss)$ integral and the fitted
 parameters can be quantified by 
calculating
the difference between the full--BEC and intra--W BEC predictions 
from MC simulation
scaled 
by
the total uncertainty of the measurement from the data. 
The values obtained for 
$J(\lss)$  and the ``strength''
 parameters 
 are 2.2, 2.3, 2.0 and 1.8 for
$J(\lss)$, $\Lambda$, $\Lambda'$ and $\Lambda_d$ respectively,
indicating that 
the powers of all these variables 
are comparable, and both the $D(\lss)$ distributions and the $J(\lss)$ 
integral 
can be used as sensitive tests for establishing inter--WW BEC. For the 
purposes of setting a limit on the amount of inter--WW BEC to be 
considered
as a systematic
  uncertainty 
 for 
  the W mass measurement, the measured data value
plus one standard deviation can be taken as a bound on the fraction
of the full \pythia\ model prediction consistent with the data. This
fraction is 77\% of the full \pythia\ inter--WW prediction for the 
 $\Lambda$ parameter
and 44\% for the $J(\lss)$ integral. It should also be noted that the 
$D$
and $J$ variables are sensitive to correlations other 
than BEC, but these
effects are expected to cancel (assuming charge independence) in the
$\delta\rho$, $\Delta_I$ and $d$  functions. Therefore, exploring both the 
$D(\lss)$ and e.g. the 
$\delta\rho$
distributions
(both of which disfavour inter--WW BEC) gives complementary information.
 
\section{Conclusions} 

The full 
sample of high-energy $\epem \to \WW$ 
events collected by the
OPAL detector has been studied 
 to look for evidence of 
Bose-Einstein correlations between like-sign hadron pairs from  different
W bosons,
using 
dedicated
test  variables and
distributions.
The   model for BEC effects as implemented in \pythia\
has been used to demonstrate  the sensitivity.
Within the data statistics available, no 
 inter--WW BEC
effects have been observed. 
  Inter--WW BEC effects of  the size predicted by \pythia\ are disfavoured. 
However, the limited data statistics do not permit them to be completely 
excluded. 
On the basis of the {\pythia} model, 
 the $D(\lss)$ 
 distribution  and the integral 
integral of the $\Delta\rho(\lss)$ distribution, $J(\lss)$, 
  are
found to 
be 
the most sensitive to inter--WW BEC
 out of all variables studied.
 The measured value of the
 $J(\lss)$ 
 is 
 found 
to be
  2.2 standard deviations
 below
 the value 
expected from 
the 
 \pythia\ full--BEC scenario.
A combination of the data from all four LEP experiments will be required
to 
draw a firm conclusion 
on the existence or absence of inter--WW BEC effects.

\bigskip
\bigskip

\nopar
{\it Acknowledgements}:
\bigskip

\nopar
We particularly wish to thank the SL Division for the efficient operation
of the LEP accelerator at all energies
 and for their close cooperation with
our experimental group.  In addition to the support staff at our own
institutions we are pleased to acknowledge the  \\
Department of Energy, USA, \\
National Science Foundation, USA, \\
Particle Physics and Astronomy Research Council, UK, \\
Natural Sciences and Engineering Research Council, Canada, \\
Israel Science Foundation, administered by the Israel
Academy of Science and Humanities, \\
Benoziyo Center for High Energy Physics,\\
Japanese Ministry of Education, Culture, Sports, Science and
Technology (MEXT) and a grant under the MEXT International
Science Research Program,\\
Japanese Society for the Promotion of Science (JSPS),\\
German Israeli Bi-national Science Foundation (GIF), \\
Bundesministerium f\"ur Bildung und Forschung, Germany, \\
National Research Council of Canada, \\
Hungarian Foundation for Scientific Research, OTKA T-038240, 
and T-042864,\\
NWO/NATO Fund for Scientific Research, the Netherlands.\\

\newpage

\begin{table}[h]
\caption{
The 
 integral $J\equiv \int_0^{Q_{\rm max}} \Delta \rho(Q) \mathrm{d}Q$
for $Q_{\rm max}=0.48$~GeV 
and results of 
fits 
 of Eq.~(\ref{fit}) to
$D(Q)$, $D'(Q)$ of like-sign pairs and 
 $d(Q)$.
For the data, the first uncertainty is
statistical and the second systematic, whilst 
only statistical uncertainties are given
for the Monte Carlo 
values.
}
 \begin{center}
\small
\begin{tabular}{cccccc}
\hline Variable & Parameter  & Data & no BEC & intra--W BEC  & full BEC\\
 \hline  
\smallskip
 $J(\lss)$ & & $0.17\pm 0.26\pm 0.23$ & $ 0.16 \pm 0.08$
                & $0.18 \pm 0.08$ &  $0.95 \pm 0.09$ 
\\
 \hline  
\smallskip
 $D(\lss)$ & $\Lambda$ & $0.063 \pm 0.036 \pm 0.038$ & $ 0.006 \pm 0.011$
                & $0.023 \pm 0.011$ &  $0.143 \pm 0.012$ 
\\
\smallskip
 &  $R$~(GeV)   &
\multicolumn{3}{c}{fixed at the value $R^{\, \rm full\: BEC}=0.277$} 
       & $0.277\pm0.027$\\
\smallskip
 &  $N$          &  $0.987\pm 0.013 \pm 0.018$ 
                  & $1.003\pm 0.004$
                   & $0.997\pm 0.004$ & $0.978\pm 0.003$\\
\smallskip
 &  $\delta$ &  $0.001\pm 0.009 \pm 0.018$ 
                 &   $\!\!\!\! -0.003\pm 0.003$ 
                  &  $\!\!\!\! -0.0003\pm 0.0030$ &  $0.001\pm 0.003$\\
\hline
\smallskip
  $D'(\lss)$  &  $\Lambda'$ & $0.059 \pm 0.039 \pm 0.047$ 
                 &  ---  & $0.017\pm 0.016$  & $0.140\pm 0.017$ \\
\smallskip
 &  $R'$~(GeV) & 
\multicolumn{3}{c}{fixed at the value $R'^{\, \rm full\: BEC}=0.249$} 
& $0.249\pm0.075$\\
\smallskip
 &  $N'$       & $0.986 \pm 0.012 \pm 0.012$ 
                  & ---
                   & $0.995\pm 0.006$ & $0.979\pm 0.005$\\
\smallskip
               &  $\delta'$ & $0.003 \pm 0.009 \pm 0.019$ 
                 &   --- 
                  &  $0.003\pm 0.004$ &  $0.014\pm 0.004$\\
\hline 
\smallskip
 $d$ &$\Lambda_d$ & $0.001 \pm 0.052 \pm 0.050$ 
           & $\!\!\!\! -0.0002\pm0.0171$ & $0.018\pm0.017$  & 
$0.151\pm0.018$\\
\smallskip
  &  $R_d$~(GeV) &    
\multicolumn{3}{c}{fixed at the value $R_d^{\, \rm full\: BEC}=0.158$} 
  & $0.158\pm0.028$\\
\smallskip
 &  $N_d$       & $1.004 \pm 0.014 \pm 0.038$ 
                  & $0.999\pm 0.005$
                   & $0.995\pm 0.005$ & $0.986\pm 0.005$\\
\smallskip
 &  $\delta_d$  & $\!\!\!\! -0.004\pm 0.011 \pm 0.019$ 
                 &   $0.001\pm 0.003$ 
                  &  $0.003\pm 0.004$ &  $0.006\pm 0.004$\\
\hline 
\end{tabular}
 \end{center}
\label{tab-fit}
\end{table}

\begin{table}[h]
 \vspace*{-.65cm}
\caption{
Contributions to the systematic uncertainty on 
 $\Lambda$,
 $\Lambda'$ and $\Lambda_d$ 
 parameters 
  in fits of
respectively,  $D(Q)$, $D'(Q)$ and $d(Q)$ 
 with 
Eq.~(\ref{fit}). 
}
\begin{center}
   \vspace*{-.45cm}
\small
\begin{tabular}{lccc}
\hline
 Source & Uncertainty on $\Lambda$   &  Uncertainty on $\Lambda'$\
& Uncertainty on $\Lambda_d$\
  \\
\hline  
  Distance of track closest approach &  0.016 & 0.014 & 0.007\\
  Specific energy loss &  0.006 & 0.013 & 0.011\\
  \WWqqln\ event bakground & 0.008& 0.011& 0.027\\
  \WWqqqq\ event likelihood weight & 0.003 & 0.002  & 0.004\\
4-jet scaling & 0.017 &  0.019  & 0.003\\
  Maximal mixing angle of 50 mrad & 0.008 & 0.012 &0.004 \\
  Maximal mixing angle  of 100 mrad & 0.003 & 0.014  & 0.004\\
  Mixing energy difference  & 0.012 & 0.005  & 0.004\\
  Fit range & 0.019 & 0.023  & 0.032\\
  $R^{\, \rm full\:\:BEC}$ errors & 0.003&0.018 & 0.008\\ 
 Removing $1+\delta \cdot Q$ from the fit&  0.013&  0.010 &0.020 \\
& & &  \\
Total systematic uncertainty & 0.038 & 0.047 & 0.050  \\
\hline 
\end{tabular}
\end{center}
\label{tab-events}
\end{table}

\newpage

\begin{figure}[t]
\center{\includegraphics[bb=30 129 575 638, width=15.3cm]
 {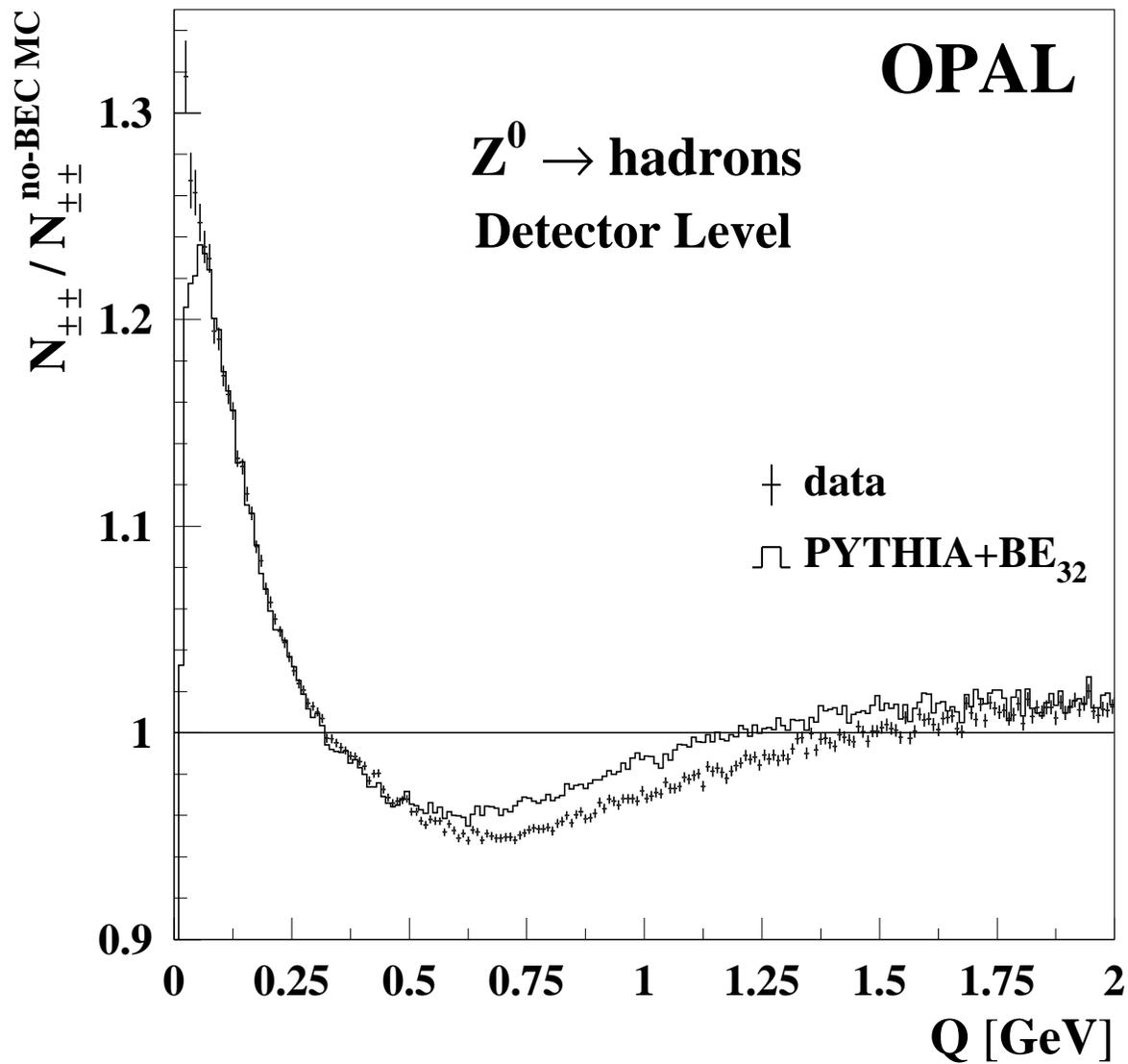}
}
{\caption{
Ratio $N_{\lss}/N_{\lss}^{\rm no-BEC\: MC}$  of the number 
$N_{\lss}$ of like-sign pairs in
\Zz\
data at 91~GeV (points) and  
\pythia\ BEC MC (histogram) to the number $N
^{\rm no-BEC\: MC}_{\lss}$
in 
standard OPAL MC without BEC.
The error bars show only the statistical uncertainties.
\label{Fig:tune}}}
\end{figure}

\begin{figure}[t]
\center{\includegraphics[bb=39 187 535 722,, width=15.5cm]
{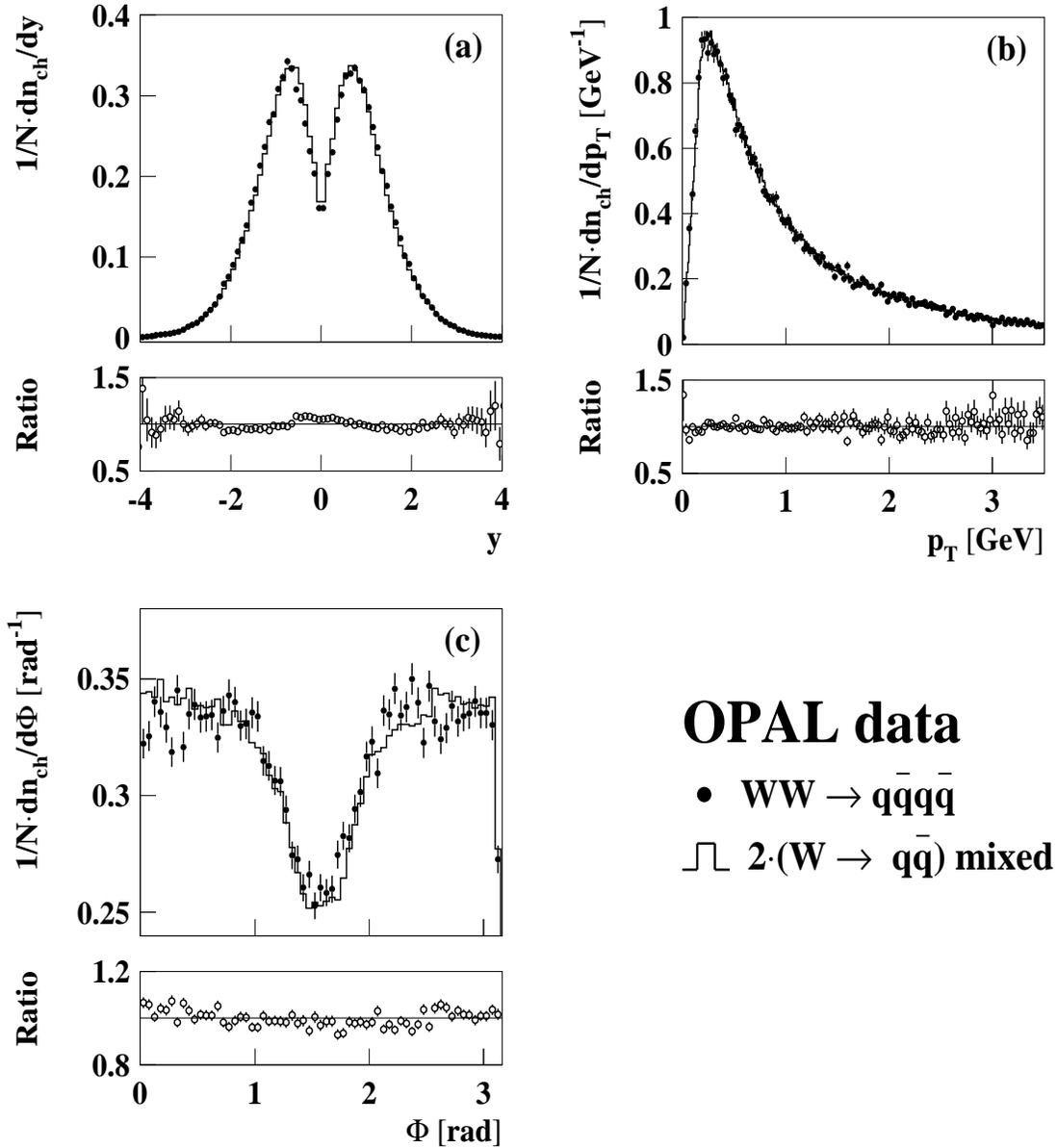}
}
{\caption{
Distributions 
of charged particles in rapidity (a),  transverse 
momentum $p_T$
(b) 
and azimuthal angle 
$\Phi$ (c)
for selected  
\WWqqqq\ data events (dots) compared 
 with 
 artificial events (histograms)
obtained by the event mixing (see text). 
The transverse  momentum is defined with respect to
the thrust axis, and azimuthal angle with respect to a plane containing 
the 
e$^-$-direction and the thrust
axis.
The lower part of each figure shows the ratio of  
 the two corresponding
 distributions. 
The error bars show only the statistical uncertainties.
\label{Fig:spectra}}}
\end{figure}

\begin{figure}[t]
\center{\includegraphics[bb=39 187 535 722, width=15.5cm]
{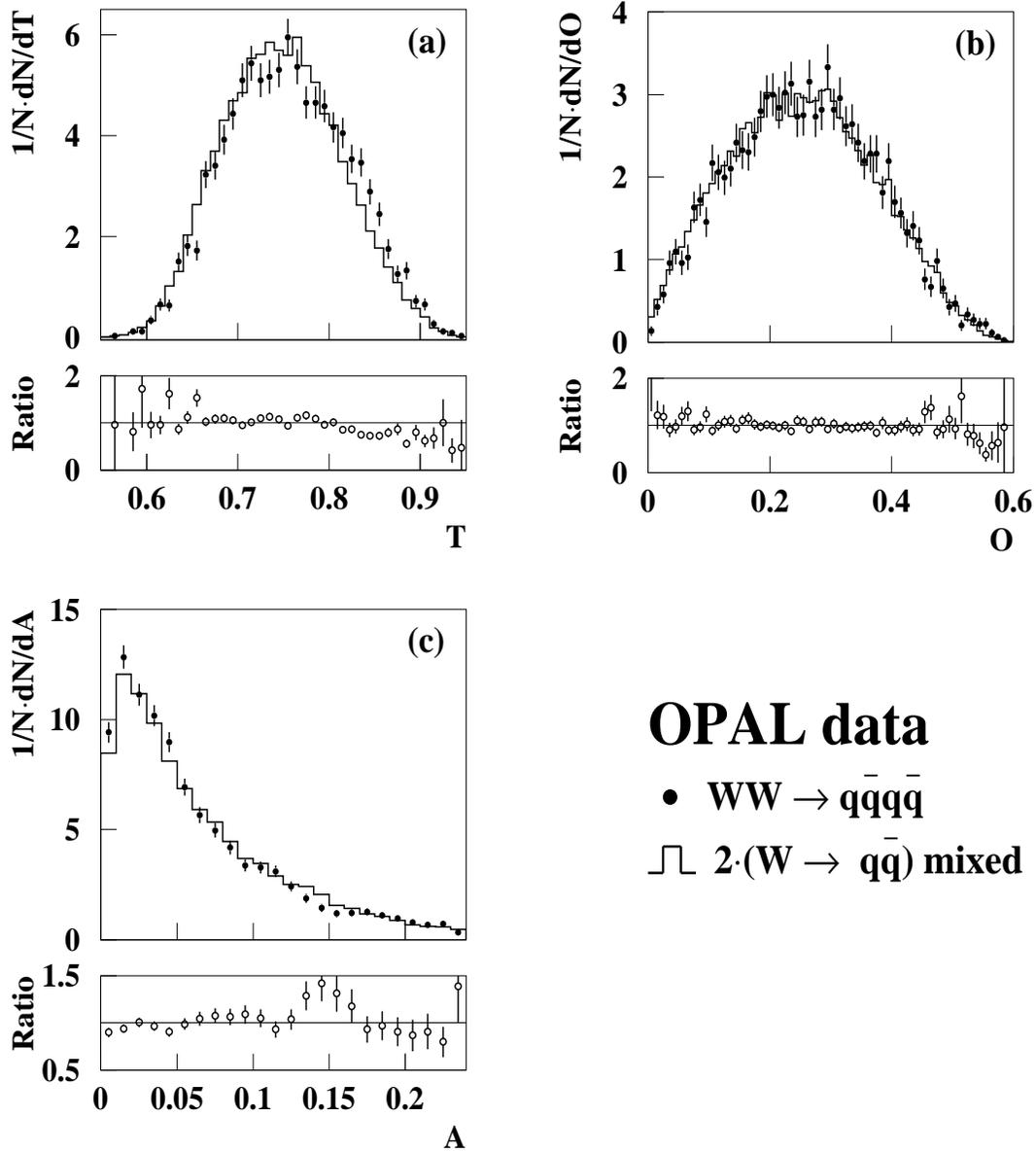}
}
{\caption{
Event shape distributions of thrust 
(a), 
oblateness (b), and aplanarity (c) 
for selected  
\WWqqqq\ data events (dots) compared 
with
 artificial events (histograms)
obtained by the event mixing (see text). 
The lower part of each figure shows the ratio of  
  the two corresponding
distributions. 
The error bars show only the statistical uncertainties.
\label{Fig:eventshapes}}}
\end{figure}

\begin{figure}[t]
\vspace*{-3.cm}
\center{\includegraphics[bb=0 120 680 660, width=12.cm]
{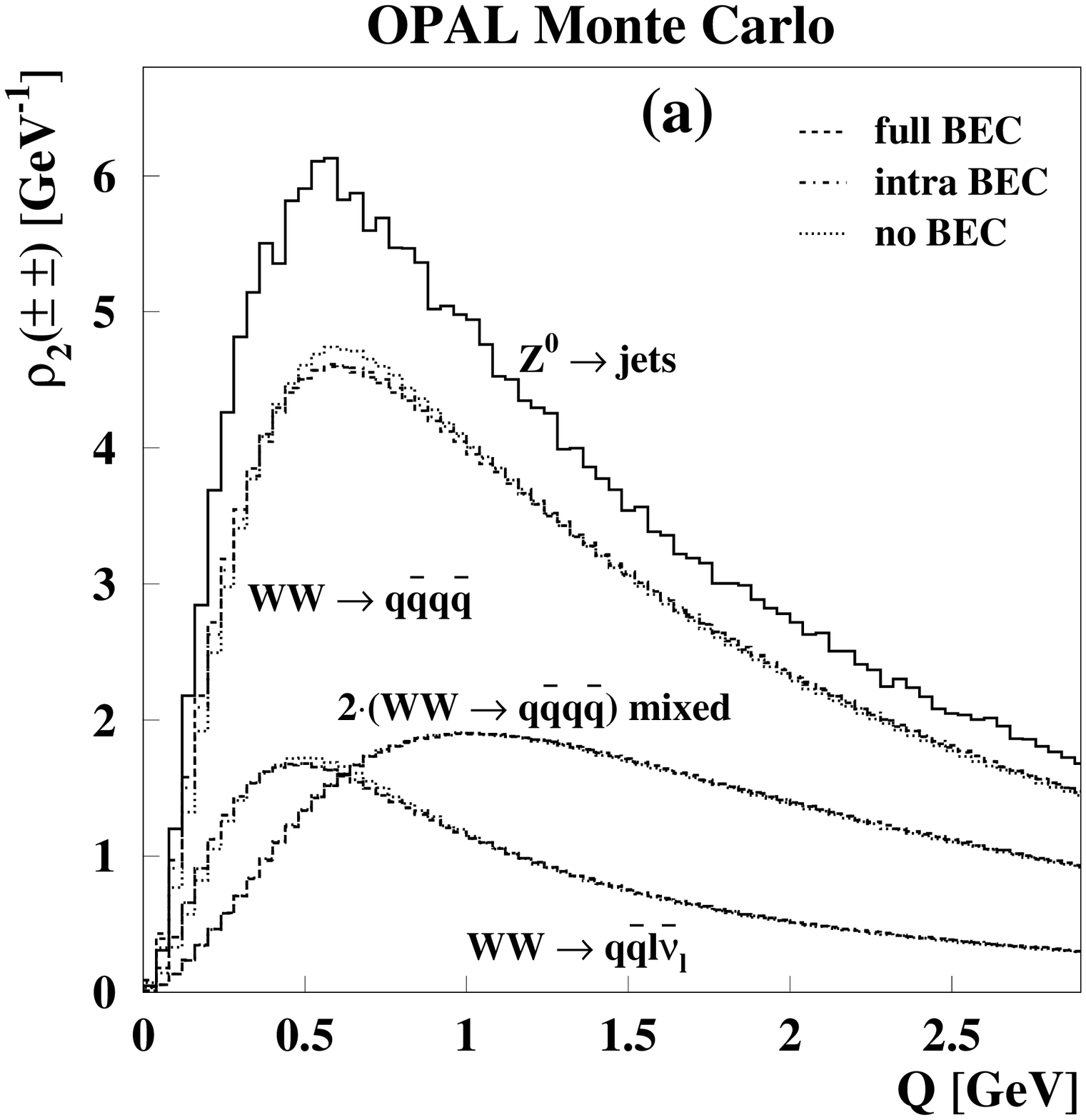}
}
\vspace*{-.45cm}
\center{\includegraphics[bb=0 120 680 660, width=12.cm]%
{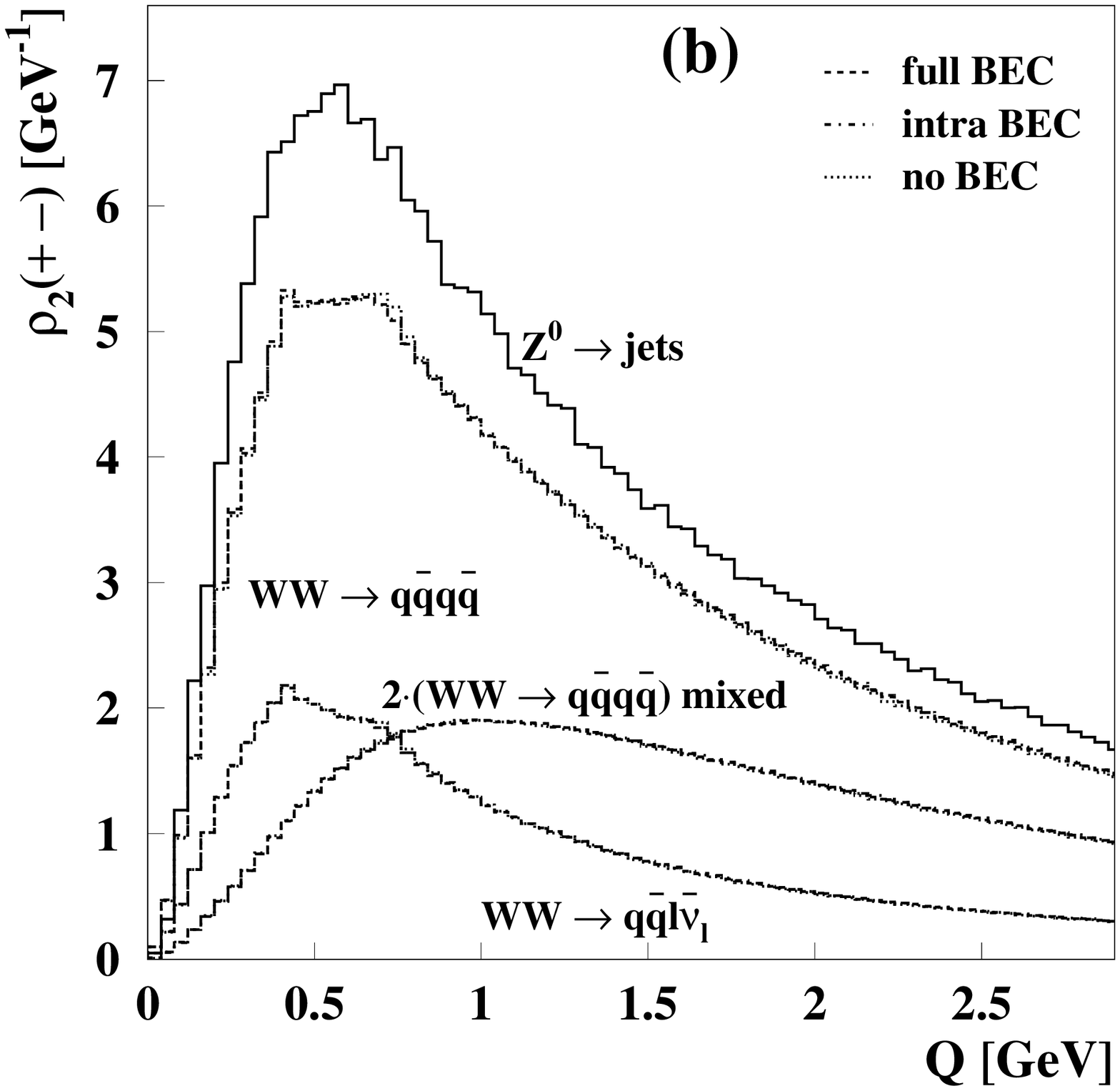}
}
{\caption{
Two-particle densities for like-sign (a) and unlike-sign (b) pairs
from MC \WWqqqq, \WWqqln\ and mixed \WWqqqq\ events 
in
different 
BEC scenarios and for the residual background from  \zg4~four jets.
\label{Fig:qdist_lus}}}
\end{figure}

\pagebreak

\begin{figure}[t]
\vspace*{-3.3cm}
\center{\includegraphics[bb=20 130 600 700, width=15.5cm]
{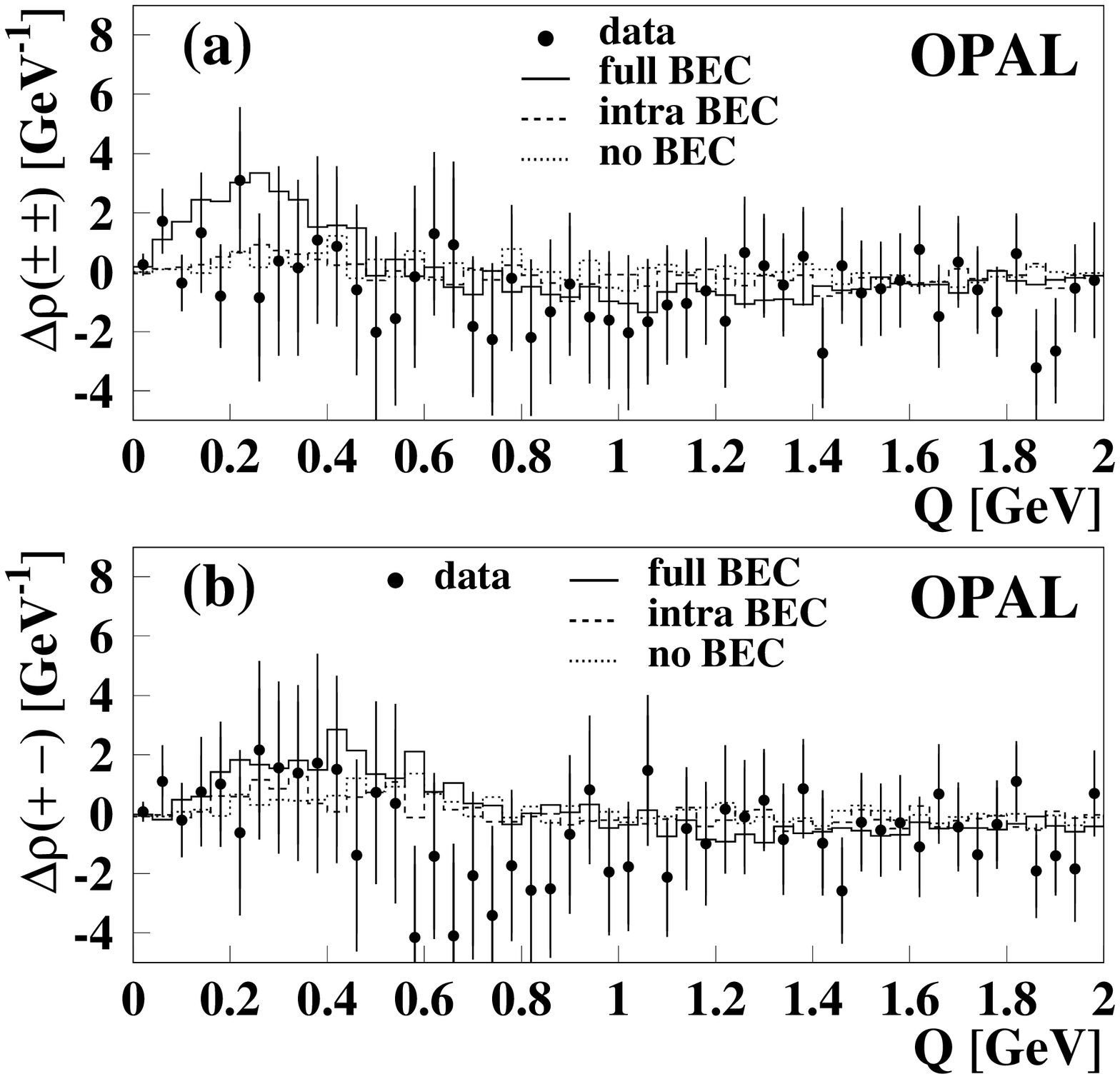}
}
\vspace*{-.2cm}
\center{\includegraphics[bb=20 350 600 641, width=15.5cm]
{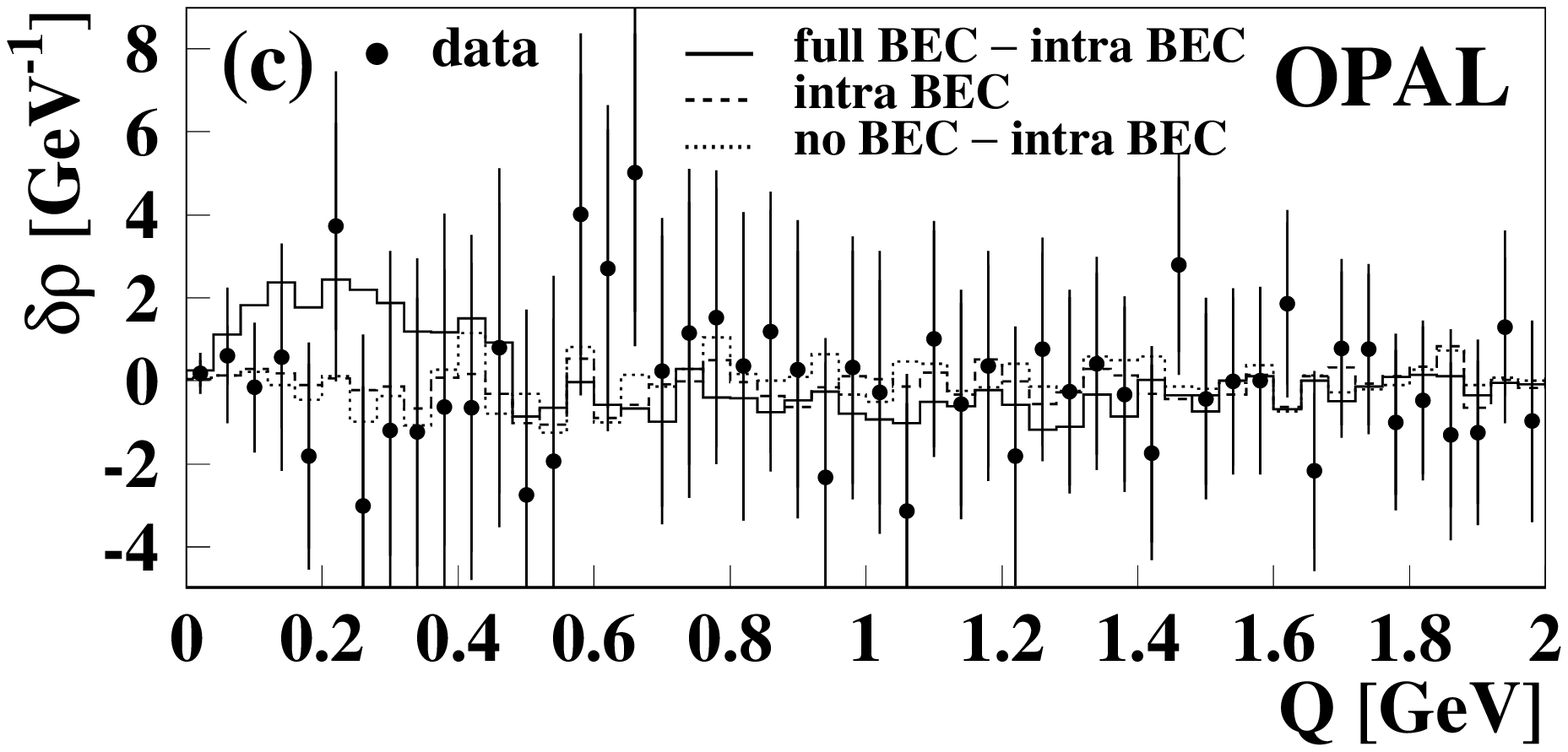}
}
\vspace*{-1.6cm}
{\caption{The $\Delta \rho(Q)$ distribution for like-sign (a) and 
unlike-sign (b) particle pairs, and the double-difference   
$\delta \rho(Q) = \Delta \rho(\lss) -  \Delta \rho(+\,-)$  (c) 
compared 
with 
different \pythia\ BEC scenarios.
\figcapb\
 In the \pythia\ predictions for the $\delta\rho$ function, the intra--W 
 scenario is used for unlike-sign pair distributions.
\label{Fig:drhos}}}
\end{figure}

\begin{figure}[t]
\vspace*{-3.3cm}
\center{\includegraphics[bb=20 380 600 700,width=15.5cm]
{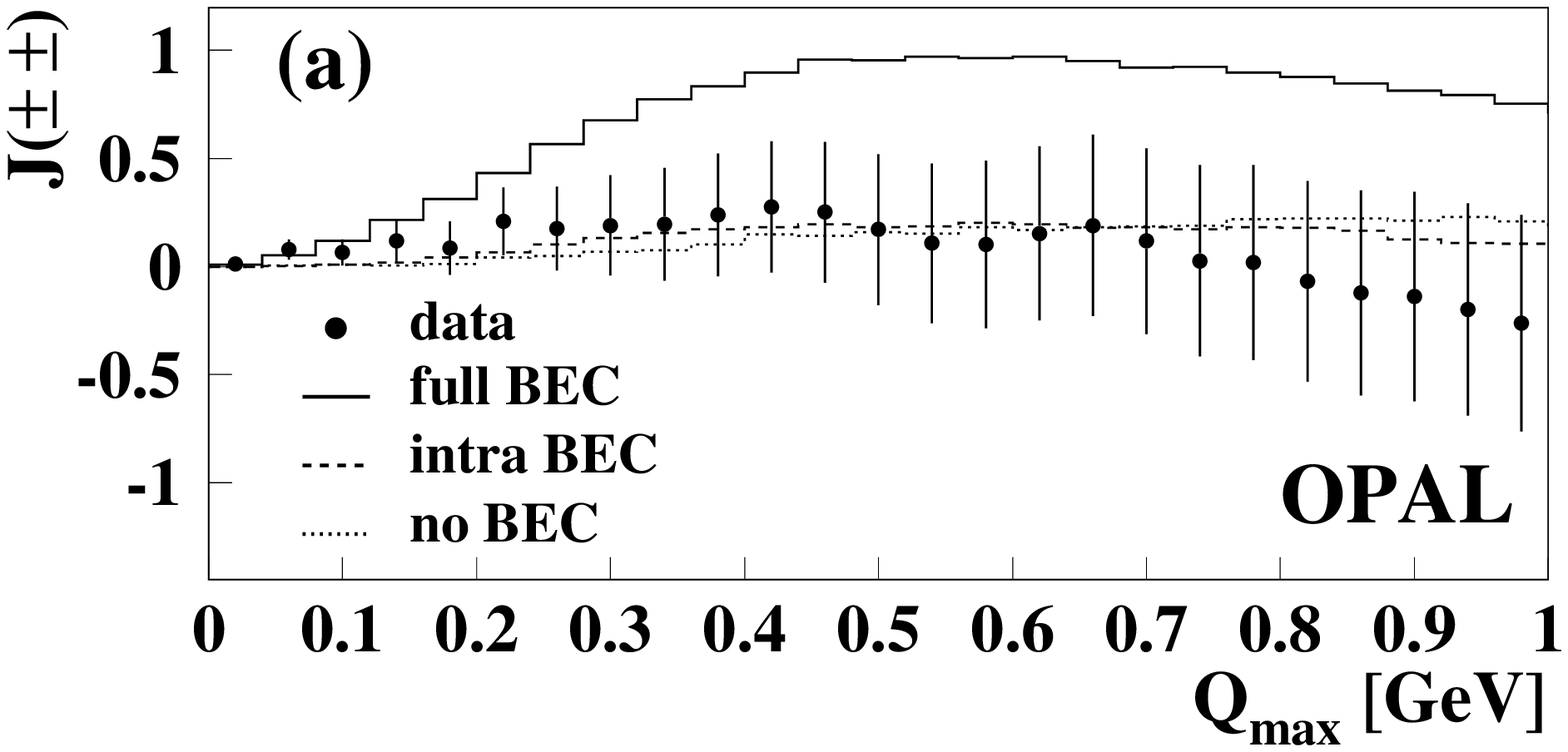}
}
\vspace*{-.2cm}
\center{\includegraphics[bb=20 380 600 670, width=15.5cm]
{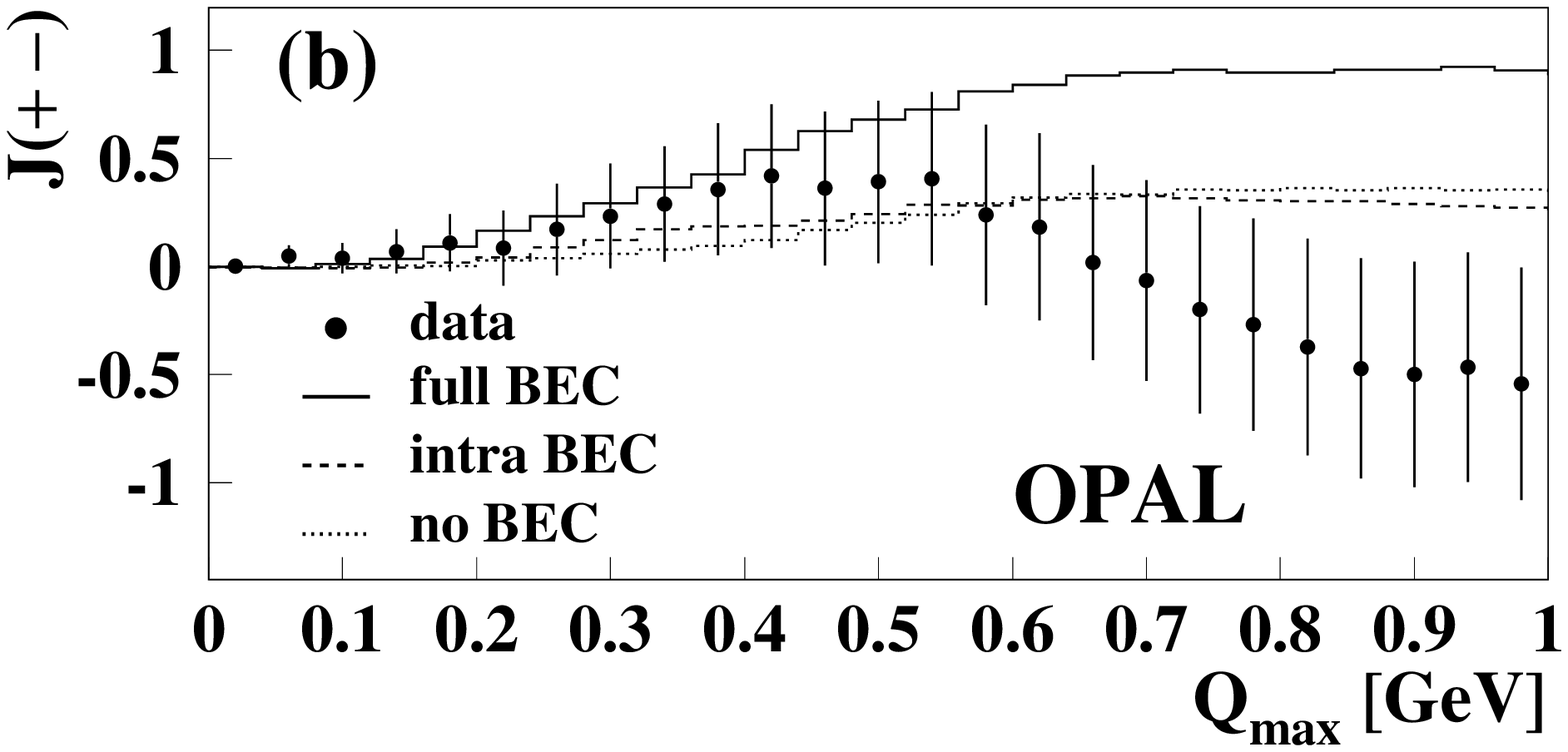}
}
{\caption{
The integral $J\equiv \int_0^{Q_{\rm max}} \Delta \rho(Q) \mathrm{d}Q$
as a function of $Q_{\rm max}$ for like-sign (a) and 
unlike-sign (b) particle pairs
compared 
 with 
different \pythia\ BEC scenarios.
 The correlated error bars show the statistical and systematic 
  uncertainties added in quadrature.
\label{Fig:int}}}
\end{figure}

\begin{figure}[t]
\vspace*{-3.3cm}
\center{\includegraphics[bb=20 120 600 700, width=15.5cm]
{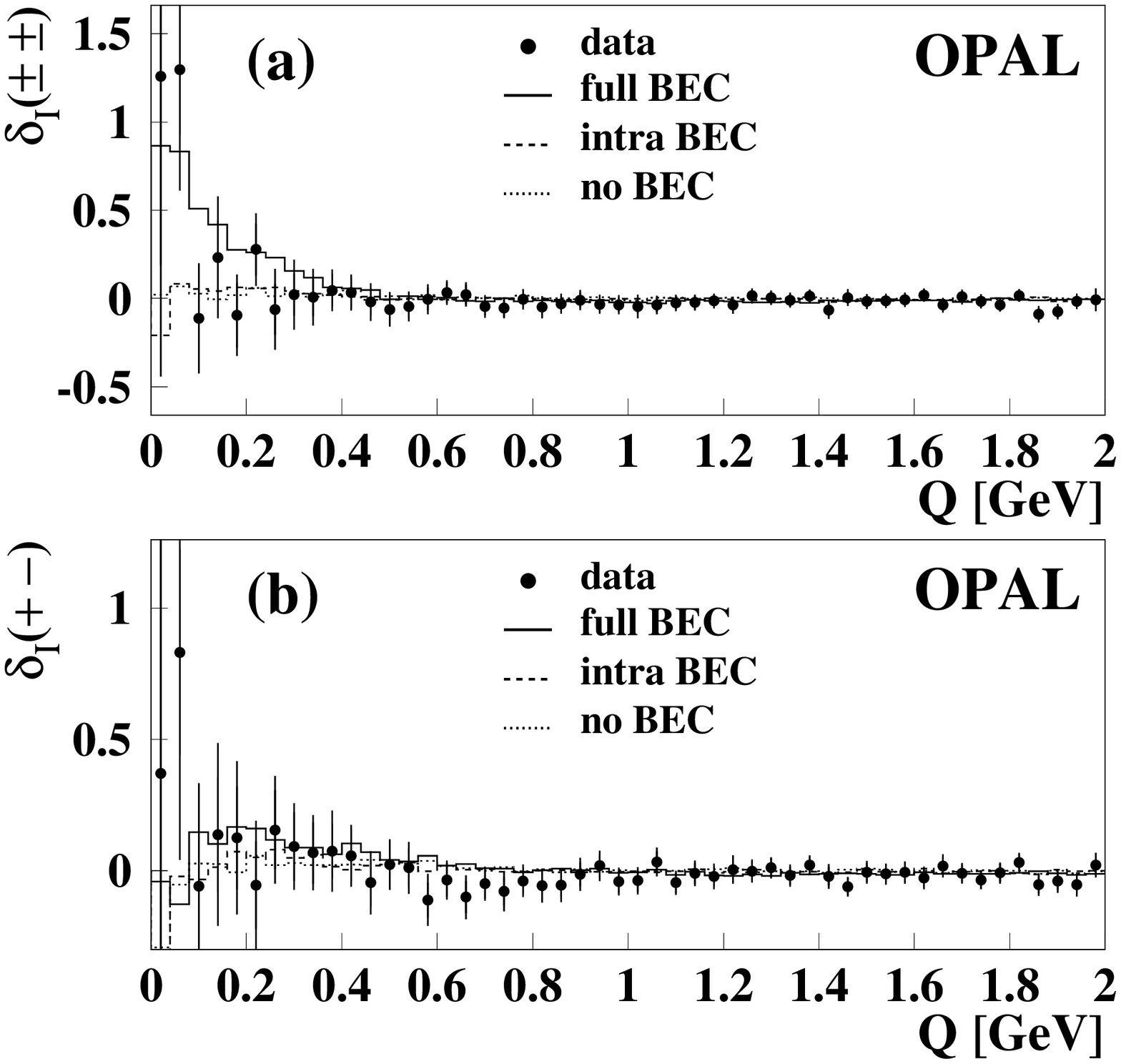}
}
\vspace*{-.2cm}
\center{\includegraphics[bb=20 350 600 641, width=15.5cm]
{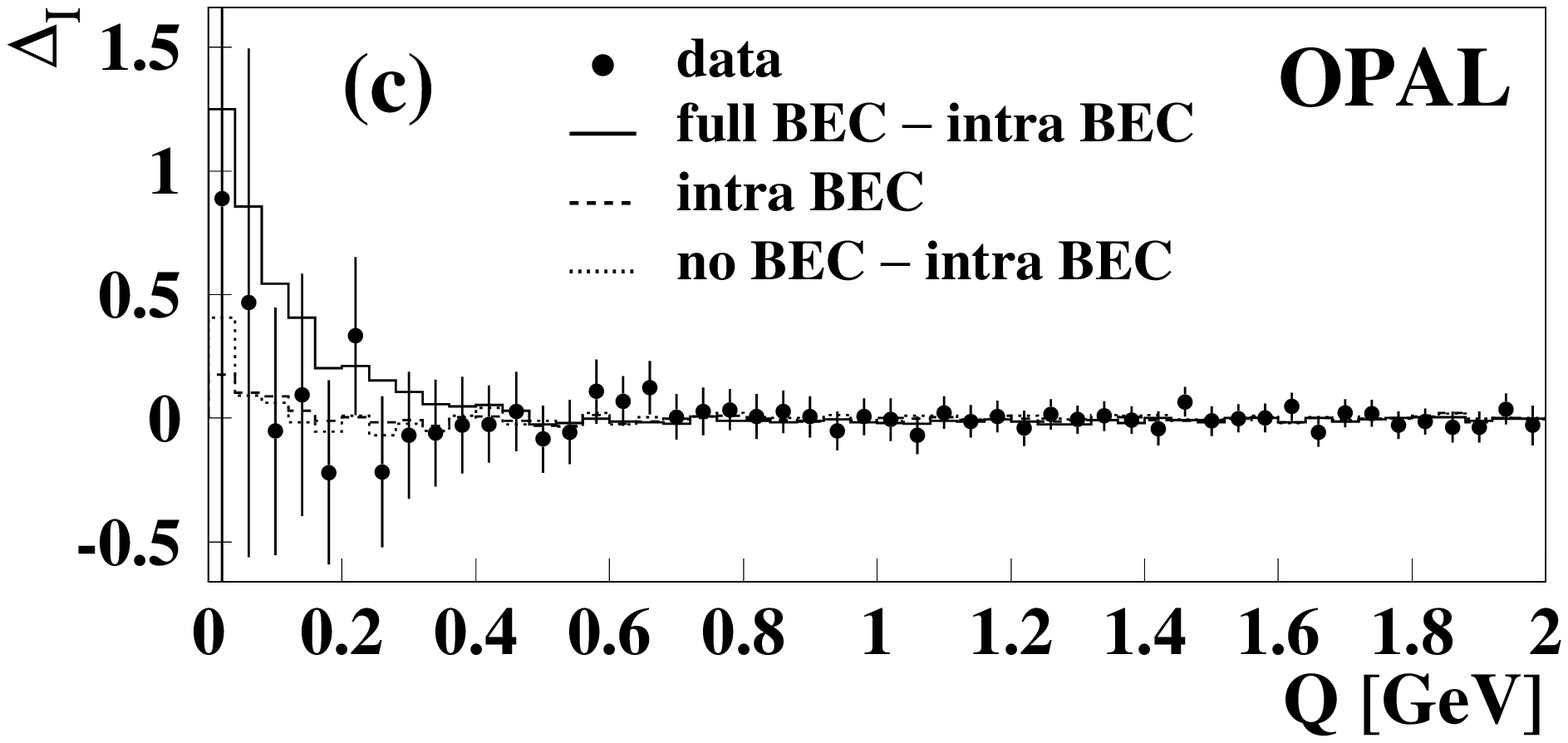}
}
\vspace*{-1.6cm}
{\caption{ The distributions of $\delta_I=\Delta \rho(Q)/\rho^{\rm
WW}_{\rm
mix}$ for like-sign (a),  unlike-sign
(b) particle pairs and 
the distribution of 
their difference 
$\Delta_I=\delta_I(\lss)-\delta_I(+\,-)$
(c),  
compared with
 different \pythia\ BEC
scenarios.  
\figcapb\
 In the \pythia\ predictions for the $\Delta_I$ function, the intra--W 
 scenario is used for unlike-sign pair distributions.
\label{Fig:deltaI}}}
\end{figure}


\begin{figure}[t]
\vspace*{-3.5cm}
\center{\includegraphics[bb=20 120 600 700, width=15.5cm]
{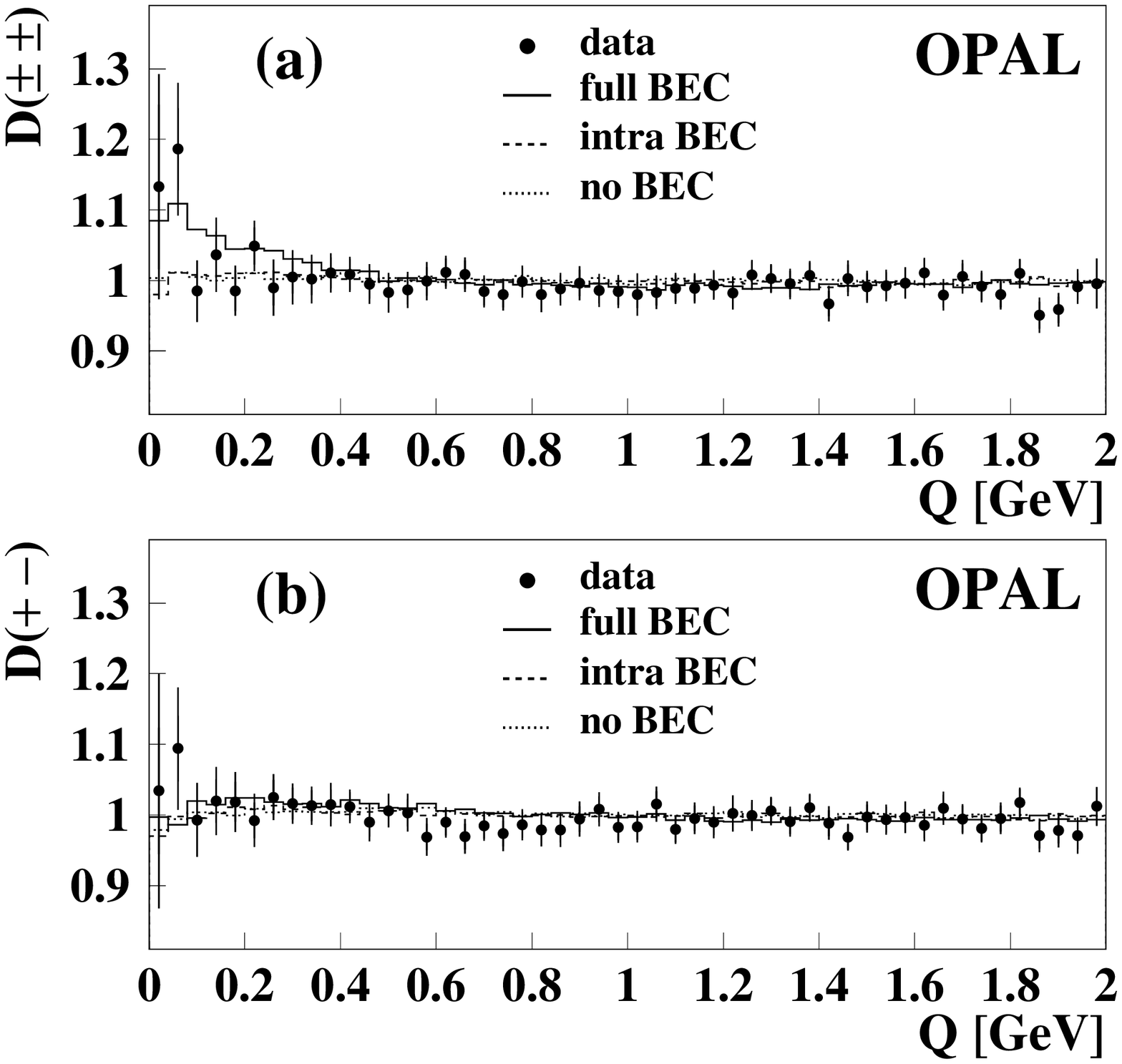}
}
\vspace*{-.4cm}
\center{\includegraphics[bb=20 350 600 641, width=15.5cm]
{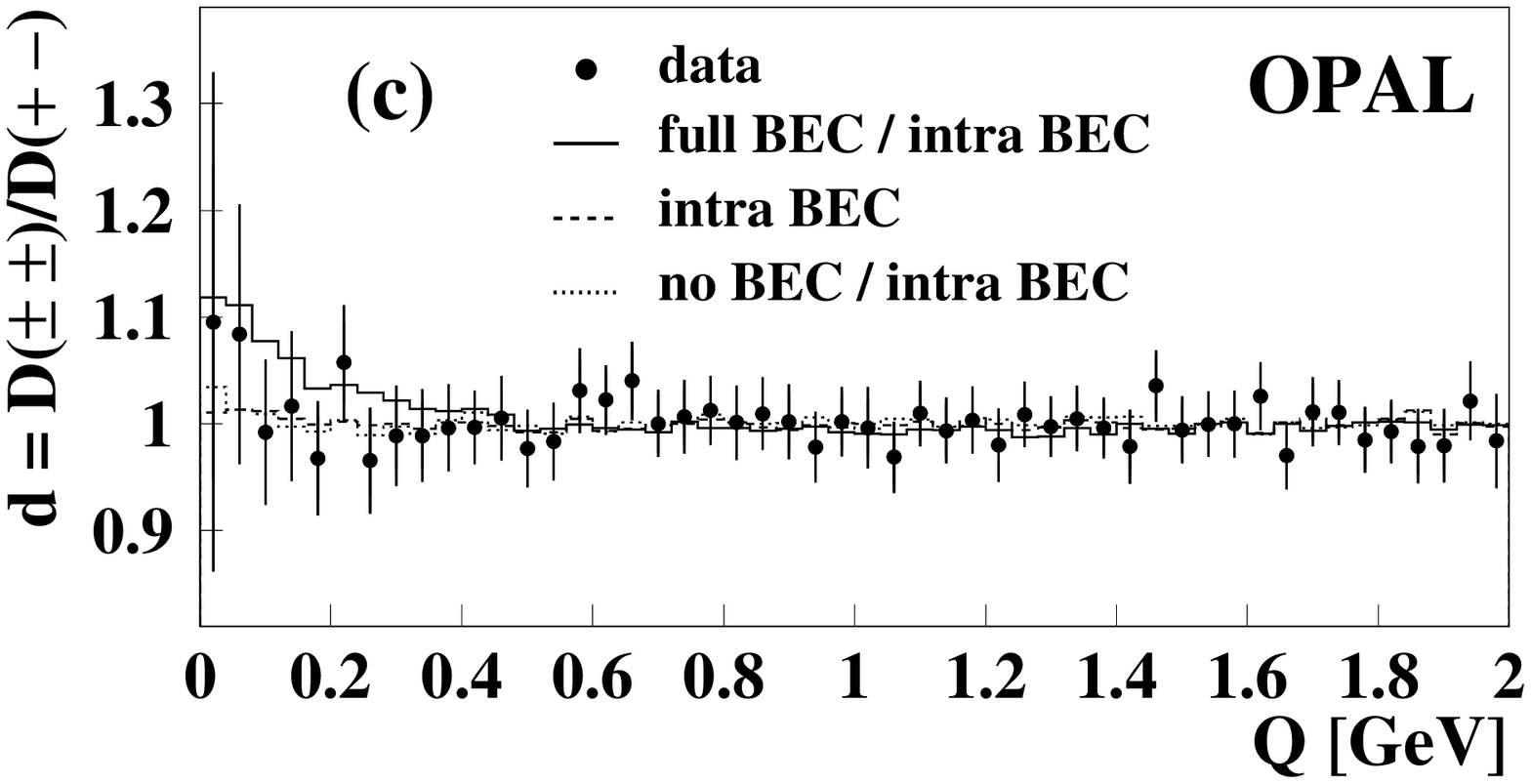}
}
\vspace*{-1.1cm}
{\caption{The $D(Q)$ spectrum for like-sign (a) and unlike-sign
(b) particle pairs, and the double-ratio $d(Q)$ distribution (c) 
compared 
with 
different \pythia\ BEC scenarios.    
\figcapb\
 In the \pythia\ predictions for the $d$-ratio, the intra--W 
 scenario is used for  unlike-sign pair distributions.
\label{Fig:D}}}
\end{figure}

\begin{figure}[t]
\center{\includegraphics[bb=20 120 600 700, width=15.5cm]
{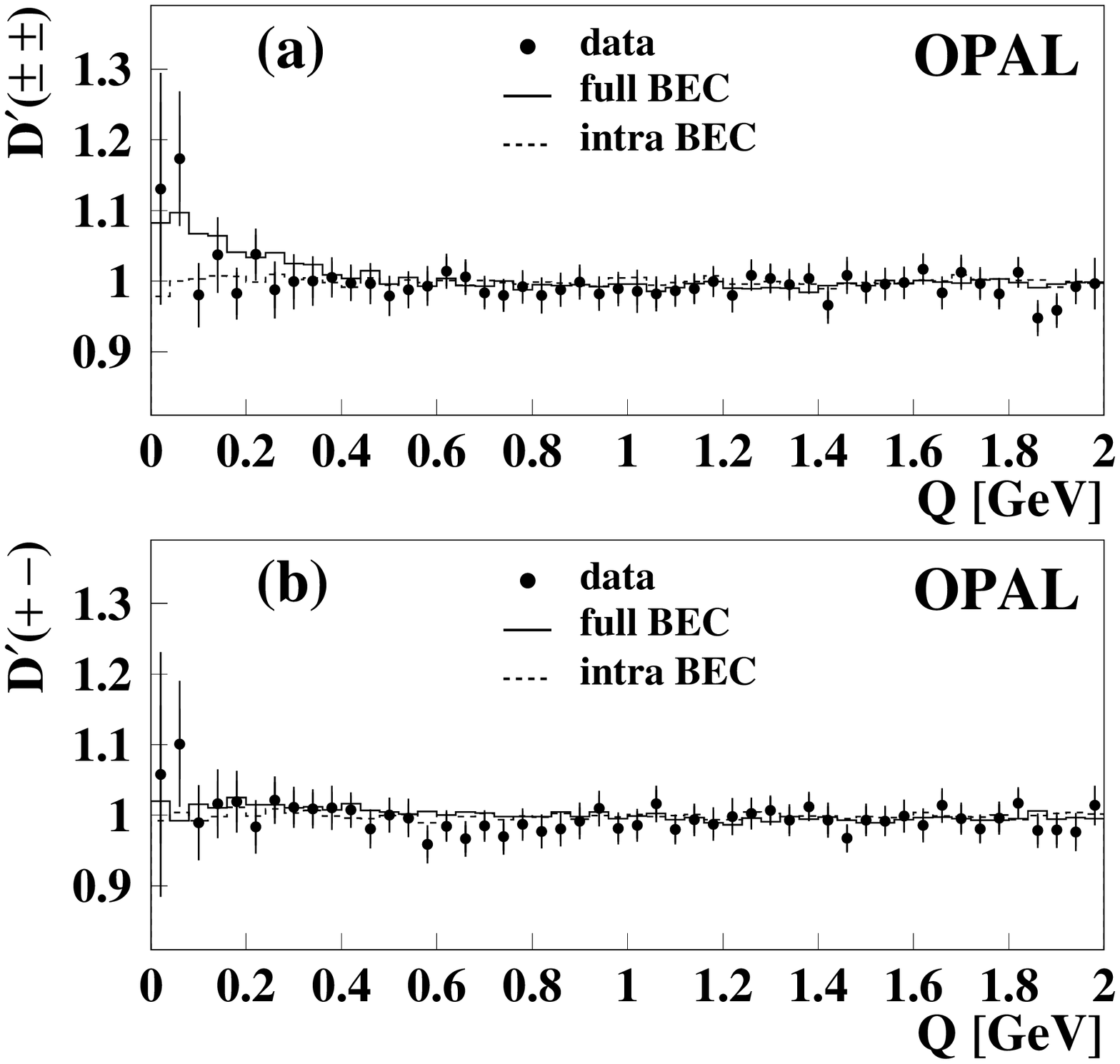}
{dps.ps}
}
{\caption{ $D'(Q)$-distributions for like-sign (a) and unlike-sign
(b) particle pairs
compared with different \pythia\ BEC
scenarios.  \figcapb
\label{Fig:Dprime}}}
\end{figure}

\begin{figure}[t]
\center{\includegraphics[bb=20 120 600 700, width=9.5cm]
{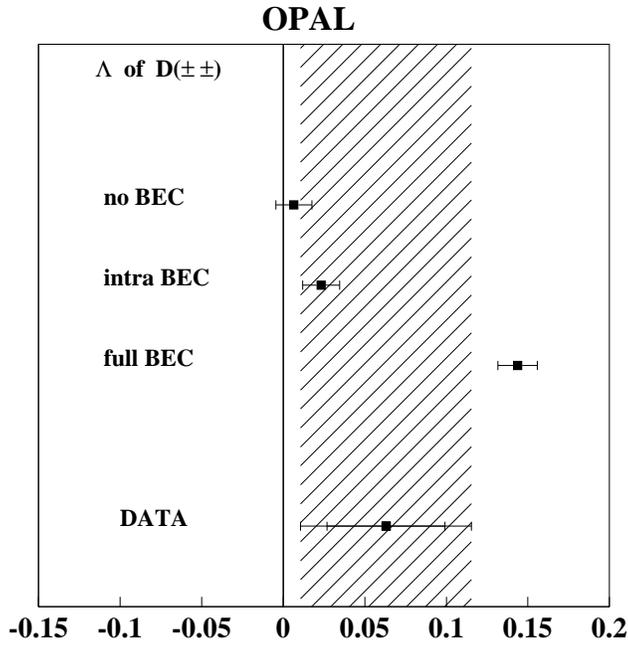}
}
\vspace*{-0.6cm}
\center{\includegraphics[bb=20 120 600 700, width=9.5cm]
{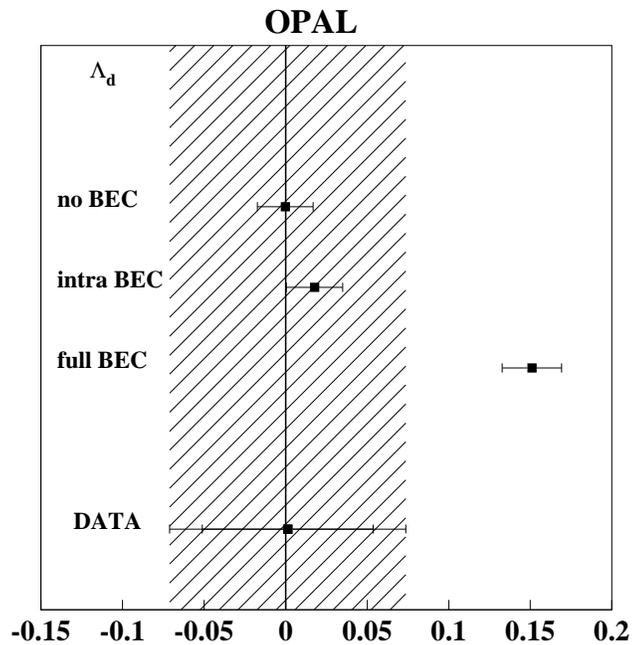}
}
{\caption{Fit results for the $\Lambda$  and  $\Lambda_d$
parameters for
data compared with different 
\pythia\ BEC scenarios.
The inner error bars for the data results and the error bars for the MC 
predictions show only the statistical uncertainties. 
The shaded 
area 
shows 
the statistical and systematic 
uncertainties 
added in quadrature.
\label{Fig:fit}}}
\end{figure}

\end{document}